\documentclass[twocolumn]{aastex61}
\pdfoutput=1 
\usepackage{amsmath,amstext}
\usepackage[T1]{fontenc}
\usepackage{apjfonts} 
\usepackage[figure,figure*]{hypcap}

\newcommand{\kms}{{\hbox {km\thinspace s$^{-1}$}}}

\newcommand{\Msun}{{\hbox {M$_\odot$}}}

\newcommand{\cmd}{{\hbox {cm$^{-2}$}}}
\newcommand{\hdo}{{\hbox {H$_{2}$O}}}
\newcommand{\nht}{{\hbox {NH$_{3}$}}}

\newcommand{\tgas}{{\hbox {$T_{\mathrm{gas}}$}}}

\newcommand{\tdust}{{\hbox {$T_{\mathrm{dust}}$}}}

\newcommand{\ohp}{{\hbox {OH$^+$}}}
\newcommand{\hdop}{{\hbox {H$_2$O$^+$}}}
\newcommand{\htop}{{\hbox {H$_{3}$O$^{+}$}}}

\shorttitle{Outflowing OH$^+$ in Markarian 231}
\shortauthors{Gonz\'alez-Alfonso et al.}

\begin{document}

\title{Outflowing OH$^+$ in Markarian 231: the ionization rate of the
  molecular gas} 

\author{E. Gonz\'alez-Alfonso}
\affiliation{Universidad de Alcal\'a, Departamento de F\'{\i}sica
     y Matem\'aticas, Campus Universitario, E-28871 Alcal\'a de Henares,
     Madrid, Spain}
\affiliation{Harvard-Smithsonian Center for Astrophysics, 60 Garden Street,
  Cambridge, MA 02138, USA}
\author{J. Fischer}\thanks{Current address:
  George Mason University, Department of Physics \& Astronomy, MS 3F3, 4400
  University Drive, Fairfax, VA 22030, USA} 
\affiliation{Naval Research Laboratory, Remote Sensing Division, 4555
     Overlook Ave SW, Washington, DC 20375, USA}
\author{S. Bruderer}
\affiliation{Max-Planck-Institute for Extraterrestrial Physics (MPE),
  Giessenbachstra{\ss}e 1, 85748 Garching, Germany}
\author{M. L. N. Ashby}
\affiliation{Harvard-Smithsonian Center for Astrophysics, 60 Garden Street,
  Cambridge, MA 02138, USA}
\author{H. A. Smith}
\affiliation{Harvard-Smithsonian Center for Astrophysics, 60 Garden Street,
  Cambridge, MA 02138, USA}
\author{S. Veilleux}
\affiliation{Department of Astronomy and Joint
Space-Science Institute, University of Maryland, College Park, MD
  20742, USA}
\author{H.~S.~P. M\"uller}
\affiliation{I.~Physikalisches Institut, Universit{\"a}t zu K{\"o}ln,
Z{\"u}lpicher Str. 77, 50937 K{\"o}ln, Germany}
\author{K. P. Stewart}
\affiliation{Naval Research Laboratory, Remote Sensing Division, 4555
     Overlook Ave SW, Washington, DC 20375, USA}
\author{E. Sturm}
\affiliation{Max-Planck-Institute for Extraterrestrial Physics (MPE),
  Giessenbachstra{\ss}e 1, 85748 Garching, Germany}

\begin{abstract}
The oxygen-bearing molecular ions OH$^+$, H$_2$O$^+$, and H$_3$O$^+$ are key
species that probe the ionization rate of (partially) molecular gas that is
ionized by X-rays and cosmic rays permeating the interstellar medium.
We report {\it Herschel} far-infrared and submillimeter spectroscopic
  observations of OH$^+$ in Mrk~231, showing both ground-state P-Cygni
  profiles, and excited line profiles with blueshifted absorption wings
  extending up to $\approx1000$\,km\,s$^{-1}$. In addition, OH$^+$ probes an
  excited component peaking at central velocities, likely arising from the
  torus probed by the OH centimeter-wave megamaser. 
Four lines of H$_2$O$^+$ are also detected at systemic
velocities, but H$_3$O$^+$ is undetected. 
Based on our earlier OH studies, we estimate an abundance
ratio of $\mathrm{OH/OH^+}\sim5-10$ for the outflowing components and
$\approx20$ for the torus, and an \ohp\ abundance relative to H nuclei of
$\gtrsim10^{-7}$. We also find high \ohp/\hdop\ and
\ohp/\htop\ ratios, both are $\gtrsim4$ in the torus and $\gtrsim10-20$
in the outflowing gas components. Chemical models indicate that 
these high \ohp\ abundances relative to OH, \hdop, and 
\htop\ are characteristic of gas with a high ionization rate per unit
density, $\zeta/n_{\mathrm H}\sim(1-5)\times10^{-17}$\,cm$^3$\,s$^{-1}$ and 
$\sim(1-2)\times10^{-16}$\,cm$^3$\,s$^{-1}$ for the above components,
respectively, an ionization rate of $\zeta\sim(0.5-2)\times10^{-12}$ s$^{-1}$,
and a low molecular fraction, $f_{\mathrm H_2}\sim0.25$. 
X-rays appear to be unable to explain the inferred ionization rate, and thus
we suggest that low-energy ($10-400$ MeV) cosmic-rays are primarily
responsible for the ionization with
$\dot{M}_{\mathrm{CR}}\sim0.01$\,M$_{\odot}$\,yr$^{-1}$ 
and $\dot{E}_{\mathrm{CR}}\sim10^{44}$ erg s$^{-1}$, the latter corresponding to
$\sim1$\% of the AGN luminosity and similar to the energetics of the molecular
outflow. We suggest that cosmic-rays accelerated in the forward shock
associated with the molecular outflow are responsible for the ionization, as
they diffuse through the outflowing molecular phase downstream. 
\end{abstract}


\section{Introduction}

Massive and powerful molecular outflows in galaxies have been detected and
reported in a number of species, including OH in the far-infrared
\citep[][hereafter GA14 and
  GA17]{fis10,stu11,spo13,vei13a,sto16,gon14,gon17} 
and CO, HCN, HCO$^+$, and CS at (sub)millimeter wavelengths
\citep[e.g.][]{sak09,fer10,fer15,ala11,ala15,cic12,cic14,aal12,aal15,gar15,lin16,vei17}. 
Each species gives specific information on the physical and chemical
conditions of the outflowing molecular gas, as a result of the different
excitation mechanisms, varying sensitivity to the physical conditions
(radiation field, gas temperature, column and volume densities), and
the diversity of chemical reations that produce and destroy the molecules. In
particular, the chemistry of the molecular gas is strongly influenced by the
extreme environments in galaxy nuclei where the gas is subject to strong
shocks, hard radiation fields (X-rays, UV), and cosmic rays (CRs). These
physical processes can be studied via the chemical pathways they cause, 
as the molecular abundances and observed line strengths are constrained by the
primary effects that X-rays and CRs have on the thermal state, partial
ionization, and dissociation of the exposed molecular gas.  

The transient OH$^+$, with its rotational transitions in the far-IR and
submillimeter domains, is one of the most suitable molecular species 
to trace the ionization rate of partially molecular 
gas and to constrain the sources of that ionization. 
CRs and X-rays are able to penetrate large columns of gas 
ionizing H and H$_2$ \citep[e.g.,][]{her73,mal96,mei11,bay11} at a rate
$\zeta$; a relatively simple chain of ion-neutral reactions then leads to the
formation of OH$^+$, which is rapidly destroyed either by reacting with H$_2$
or by dissociative recombination, at a rate proportional to $n_{\mathrm{H}}$.
Therefore, the equilibrium abundance of OH$^+$ is sensitive to both
$\zeta/n_{\mathrm{H}}$ and the molecular fraction $f_{\mathrm{H2}}$,
constraining the fluxes of CRs and X-rays that permeate the molecular gas.

\cite{gon13} (hereafter GA13) reported on and analyzed \emph{Herschel}/PACS
\citep{pil10,pog10} detections of excited O-bearing molecular ions (\ohp,
\hdop, \htop) in both NGC~4418 and Arp~220.
The detection of excited lines of \ohp, \hdop, and \htop\ in
both galaxies allowed us to derive estimates of the ionization rates per
hydrogen nucleus density ($\zeta/n_{\mathrm{H}}$), and to 
place a lower limit on the ionization rate
$\zeta$ of several hundreds times the typical rate inferred in the Milky Way.
In Arp~220, the ground-state \ohp\ lines show P-Cygni profiles
\citep{ran11} indicative of outflowing gas, and appear to arise in the more
  spatially extended disk where the 
  ionization rate is significantly lower than in the nucleus \citep{tak16}.

In the present study, we extend our work on O-bearing molecular ions
to Mrk~231, the closest QSO, the most luminous ultraluminous infrared galaxy
(ULIRG) in the local ($z<0.1$) Universe, and the most extensively studied
extragalactic molecular outflow 
\citep[][GA14, GA17]{fis10,fer10,fer15,stu11,cic12,aal12,aal15,lin16}.
The {\it Herschel}/PACS observations of Mrk~231 show prominent absorption in
the line wings of \ohp\ up to velocities of $\approx1000$ \kms, together with
very excited absorption at systemic velocities. 
We analyze the observations of \ohp\ in combination with those of
\hdop\ and \htop, in the framework of the OH observations and radiative
tranfer models reported in GA17. 
We also include in the analysis the {\it Herschel}/SPIRE observations of
the ground-state lines of OH$^+$ and \hdop, which show prominent emission
features \citep{wer10}. 
The inferred molecular abundances are further interpreted in terms of chemical
models, which enable us to constrain
$\zeta/n_{\mathrm{H}}$ and $f_{\mathrm{H2}}$ for both the quiescent and
outflowing components, as well as the ionization rate $\zeta$ upon
conservative assumptions on the gas density. We then derive the energetics
associated with low-energy CRs, which are most likely responsible for the 
ionization, and link these CRs to the forward shock associated with the
molecular outflow.
The observations are described in \S\ref{sec:obser}, the analysis is 
developed in \S\ref{sec:analysis}, and our main findings are discussed in
\S\ref{sec:discussion}. 
As in GA17, we adopt a distance to Mrk\,231 of 186\,Mpc,  
assuming a flat Universe with $H_0=71$\,km\,s$^{-1}$\,Mpc$^{-1}$ 
and $\Omega_{\mathrm{M}}=0.27$, with $z=0.04218$.

\section{Observations}
\label{sec:obser}

Most PACS data presented here were observed as part of the {\it Herschel} Open
Time (OT2) program to obtain a full high-resolution far-infrared specrum of
Mrk 231 (PI: J. Fischer); a few of the spectra presented were observed during
the science demonstration observations of the SHINING program (PI: E. Sturm). 
As described in GA14, the
spectra were observed in high spectral sampling range-mode using first and
second orders of the grating.  The velocity resolution of PACS in first order
ranges from $\approx320$ to 180 \kms\ over the wavelength
range from 105 to 190\,$\mu$m, and in second order from $\approx210$ to $110$ 
\kms\ from 52 to 98 microns.  The data reduction 
was mostly done using the PACS reduction and calibration pipeline (ipipe)
included in HIPE 14.0.1, with calibration tree version 72,
using an oversampling of 4 and fully independent channels
(an upsample parameter of 1).  Both the molecular
absorption lines and the continua are effectively 
point-like in Mrk~231, and we have thus used the point source calibrated
spectra ``c129'', produced by scaling the emission from the central
$\approx9''\times9''$ spatial pixel to the total emission from the central
$3\times3$ spaxels (``c9''), which is itself scaled according to the
point-source correction (see also GA17). The absolute flux scale is  
robust to potential pointing jitter, with continuum flux reproducibility
of $\pm15$\%. 

The {\it Herschel}/SPIRE spectrum of Mrk\,231 was observed by the Open
Time Key Project {\it Hercules} (PI: P.~van der Werf).
The detection of the \ohp\ and \hdop\ ground-state lines was reported by
\cite{wer10} using the observations carried out on OD 209 
(ObsID: 1342187893), but here we use the more sensitive
observation from the same project taken in OD 558 \citep{ros15}. 
The SPIRE spectrometer observations cover the wavelength range 
$191-671$ $\mu$m with two spatial arrays covering two bands: SSW 
($191-318$ $\mu$m) and SLW ($294-671$ $\mu$m). 
The HIPE 15.0.1 apodized spectra were downloaded from the archive.
Mrk~231 is a point
source for SPIRE FTS ($\mathrm{HPBW}\ge17''$), and therefore only the central
detector spectra were used. In HR mode, the unapodized spectral resolution is
$\mathrm{FWHM\,(km \,s^{-1})} = 1.4472\,\lambda(\mu\mathrm{m})$, but we have 
used the apodized spectra resulting in a FWHM a factor of 1.5 
higher\footnote{SPIRE Handbook,
  http://herschel.esac.esa.int/Docs/SPIRE/spire\_handbook.pdf}. 
The spectroscopic parameters used for line
identification and radiative transfer modeling were taken from the
spectral line catalogs of the CDMS \citep[\ohp\ and \hdop,][]{mul01,mul05}.
and JPL \citep[\htop,][]{pic98}; improved parameters for \ohp\ have been
  recently reported by \cite{mar16}.

The observations of the O-bearing molecular cations used in this study are
  summarized in Table~\ref{tab:obs} and  
the equivalent widths and fluxes of the PACS and SPIRE lines are listed in 
Tables~\ref{tab:fluxes} and \ref{tab:spirefluxes}, respectively.

   \begin{table}
      \caption{Obs ID of the \ohp, \hdop, and \htop\ lines in Mrk~231.}
         \label{tab:obs}
\begin{center}
          \begin{tabular}{lcc}   
            \hline
            \noalign{\smallskip}
Transition  & $\lambda_{\mathrm{rest}}$ & Obs ID   \\  
&  ($\mu$m)  &   \\
            \noalign{\smallskip}
            \hline
            \noalign{\smallskip}
\ohp\ \, $1_0-0_1$ & $329.761$ &  1342210493\\
\ohp\ \, $1_2-0_1$ & $308.488$ &  1342210493\\
\ohp\ \, $1_1-0_1$ & $290.193$ &  1342210493\\
\ohp\ \, $2_2-1_2$ & $147.768$ &  1342253535\\
\ohp\ \, $2_1-1_0$ & $148.696$ &  1342253535\\
\ohp\ \, $2_2-1_1$ & $152.369$ &  1342253536\\
\ohp\ \, $2_3-1_2$ & $152.989$ &  1342253536\\
\ohp\ \, $2_1-1_1$ & $158.437$ &  1342186811\\
\ohp\ \, $3_2-2_1$ & $101.265$ &  1342253530\\
\ohp\ \, $3_3-2_2$ & $101.698$ &  1342253530\\
\ohp\ \, $3_4-2_3$ & $101.921$ &  1342253530\\
\ohp\ \, $3_2-2_2$ & $103.909$ &  1342253530\\
\ohp\ \, $4_3-3_2$ & $76.245$ &  1342253536\\
\ohp\ \, $4_4-3_3$ & $76.399$ &  1342253536\\
\ohp\ \, $4_5-3_4$ & $76.510$ &  1342253536\\
\ohp\ \, $5_4-4_3$ & $61.168$ &  1342253532\\
\ohp\ \, $5_5-4_4$ & $61.248$ &  1342253532\\
\ohp\ \, $5_6-4_5$ & $61.315$ &  1342253532\\
$\mathrm{oH_2O^+} \, 1_{11}\frac{1}{2}-0_{00}\frac{1}{2}$ & $263.077$&1342210493  \\ 
$\mathrm{oH_2O^+} \, 1_{11}\frac{3}{2}-0_{00}\frac{1}{2}$ & $268.850$&1342210493  \\ 
$\mathrm{pH_2O^+} \, 2_{21}\frac{3}{2}-1_{10}\frac{1}{2}$ & $104.719$&1342253530  \\ 
$\mathrm{pH_2O^+} \, 2_{21}\frac{5}{2}-1_{10}\frac{3}{2}$ & $105.742$&1342253530  \\ 
$\mathrm{oH_2O^+} \, 3_{22}\frac{5}{2}-2_{11}\frac{3}{2}$ & $88.978$ &1342253539 \\ 
$\mathrm{oH_2O^+} \, 3_{22}\frac{7}{2}-2_{11}\frac{5}{2}$ & $89.590$ &1342253539 \\ 
$\mathrm{oH_2O^+} \, 3_{13}\frac{5}{2}-2_{02}\frac{3}{2}$ & $143.265$&1342253534 \\ 
$\mathrm{oH_2O^+} \, 3_{13}\frac{7}{2}-2_{02}\frac{5}{2}$ & $143.810$&1342253534 \\ 
$\mathrm{oH_2O^+} \, 4_{04}\frac{9}{2}-3_{13}\frac{7}{2}$ & $145.917$&1342253535 \\ 
$\mathrm{oH_2O^+} \, 4_{04}\frac{7}{2}-3_{13}\frac{5}{2}$ & $146.216$&1342253535 \\ 
$\mathrm{oH_3O^+} \, 4_{3}^--3_{3}^+$ & $69.538$& 1342253534 \\ 
$\mathrm{pH_3O^+} \, 4_{1}^--3_{1}^+$ & $70.684$& 1342253534 \\ 
$\mathrm{oH_3O^+} \, 4_{0}^--3_{0}^+$ & $70.827$& 1342253534 \\ 
$\mathrm{pH_3O^+} \, 3_{2}^--2_{2}^+$ & $82.274$& 1342253537 \\ 
$\mathrm{pH_3O^+} \, 3_{1}^--2_{1}^+$ & $82.868$& 1342253537  \\ 
            \noalign{\smallskip}
            \hline
         \end{tabular} 
\end{center}
   \end{table}

   \begin{table}
      \caption{Line equivalent widths and fluxes of the PACS lines}
         \label{tab:fluxes}
\begin{center}
          \begin{tabular}{lccc}   
            \hline
            \noalign{\smallskip}
            Transition  & Velocities$^{\mathrm{a}}$    & $W_{\mathrm{eq}}$$^{\mathrm{b,c}}$  & Flux$^{\mathrm{c}}$ \\ 
            & (km\,s$^{-1}$)  & (km\,s$^{-1}$)     &  (Jy\,km\,s$^{-1}$) \\
            \noalign{\smallskip}
            \hline
            \noalign{\smallskip}
\ohp\ \, $2_2-1_2$                                        & $[-300, 200]$   & $22.2$ $( 2.8)$   & $   -394$ $(   49)$   \\
                                        & $[-1000,-300]$   & $13.4$ $( 3.0)$   & $   -237$ $(   54)$   \\
\ohp\ \, $2_1-1_0$                                        & $[-300, 200]$   & $31.5$ $( 2.6)$   & $   -553$ $(   45)$   \\
                                        & $[-1000,-300]$   & $10.4$ $( 3.0)$   & $   -182$ $(   53)$   \\
\ohp\ \, $2_2-1_1$                                        & $[-300, 200]$   & $46.5$ $( 3.0)$   & $   -774$ $(   50)$   \\
                                        & $[-1000,-300]$   & $26.5$ $( 3.7)$   & $   -441$ $(   60)$   \\
\ohp\ \, $2_3-1_2$                                        & $[-300, 200]$   & $80.0$ $( 3.2)$$^{\mathrm{d}}$   & $  -1320$ $(   52)$$^{\mathrm{d}}$   \\
                                        & $[-1000,-300]$   & $49.2$ $( 3.6)$   & $   -812$ $(   60)$   \\
\ohp\ \, $2_1-1_1$                                        & $[-300, 200]$   & $14.3$ $( 0.6)$$^{\mathrm{d}}$   & $   -229$ $(    9)$$^{\mathrm{d}}$   \\
\ohp\ \, $3_2-2_1$                                        & $[-300, 200]$   & $14.1$ $( 1.3)$   & $   -419$ $(   39)$   \\
\ohp\ \, $3_3-2_2$                                        & $[-300, 200]$   & $38.4$ $( 1.3)$$^{\mathrm{d}}$   & $  -1143$ $(   38)$$^{\mathrm{d}}$   \\
                                        & $[-1000,-300]$   & $20.6$ $( 1.5)$$^{\mathrm{d}}$   & $   -615$ $(   45)$$^{\mathrm{d}}$   \\
\ohp\ \, $3_4-2_3$                                        & $[ -300, 200]$   & $44.8$ $( 1.2)$$^{\mathrm{d}}$   & $  -1334$ $(   37)$$^{\mathrm{d}}$   \\
\ohp\ \, $4_3-3_2$                                        & $[-300, 200]$   & $< 3.7$      & $>-  124 $           \\
\ohp\ \, $4_4-3_3$                                        & $[-300, 200]$   & $ 7.6$ $( 2.3)$   & $   -255$ $(   76)$   \\
\ohp\ \, $4_5-3_4$                                        & $[-300, 200]$   & $13.0$ $( 2.1)$   & $   -440$ $(   69)$   \\
\ohp\ \, $3_2-2_2$                                        & $[-300, 200]$   & $< 3.0$      & $>-   86 $           \\
\ohp\ \, $5_4-4_3$                                        & $[-300, 200]$   & $< 4.4$      & $>-  158 $           \\
\ohp\ \, $5_5-4_4$                                        & $[-300, 200]$   & $< 4.8$      & $>-  172 $           \\
\ohp\ \, $5_6-4_5$                                        & $[-300, 200]$   & $< 4.7$      & $>-  168 $           \\
$\mathrm{pH_2O^+} \, 2_{21}\frac{3}{2}-1_{10}\frac{1}{2}$ & $[ -300, 200]$   & $< 4.6$$^{\mathrm{d}}$      & $>-  134 $$^{\mathrm{d}}$           \\
$\mathrm{pH_2O^+} \, 2_{21}\frac{5}{2}-1_{10}\frac{3}{2}$ & $[ -300, 200]$   & $   4.3$ $(  1.1)$   & $   -125$ $(   32)$   \\
$\mathrm{oH_2O^+} \, 3_{22}\frac{5}{2}-2_{11}\frac{3}{2}$ & $[ -300, 200]$   & $< 3.8$      & $>-  120 $           \\
$\mathrm{oH_2O^+} \, 3_{22}\frac{7}{2}-2_{11}\frac{5}{2}$ & $[ -300, 200]$   & $< 3.9$      & $>-  120 $           \\
$\mathrm{oH_2O^+} \, 3_{13}\frac{5}{2}-2_{02}\frac{3}{2}$ & $[-300, 200]$   & $ 8.2$ $( 2.0)$   & $   -150$ $(   36)$   \\
$\mathrm{oH_2O^+} \, 3_{13}\frac{7}{2}-2_{02}\frac{5}{2}$ & $[-300, 200]$   & $ 8.4$ $( 1.6)$   & $   -154$ $(   29)$   \\
$\mathrm{oH_2O^+} \, 4_{04}\frac{9}{2}-3_{13}\frac{7}{2}$ & $[-300, 200]$   & $< 3.8$      & $>-   58 $           \\
$\mathrm{oH_2O^+} \, 4_{04}\frac{7}{2}-3_{13}\frac{5}{2}$ & $[-300, 200]$   & $< 3.2$      & $>-   48 $           \\
$\mathrm{oH_3O^+} \, 4_{3}^--3_{3}^+$                     & $[-300, 200]$   & $< 2.7$      & $>-   92 $           \\
$\mathrm{pH_3O^+} \, 4_{1}^--3_{1}^+$                     & $[-300, 200]$   & $< 3.1$      & $>-  106 $           \\
$\mathrm{oH_3O^+} \, 4_{0}^--3_{0}^+$                     & $[-300, 200]$   & $< 2.5$      & $>-   86 $           \\
$\mathrm{pH_3O^+} \, 3_{2}^--2_{2}^+$                     & $[-300, 200]$   & $< 4.2$      & $>-  138 $           \\
$\mathrm{pH_3O^+} \, 3_{1}^--2_{1}^+$                     & $[-300, 200]$   & $< 3.7$      & $>-  120 $           \\
            \noalign{\smallskip}
            \hline
         \end{tabular} 
\end{center}
\begin{list}{}{}
\item[$^{\mathrm{a}}$] Velocity range in which equivalent widths and fluxes
  are calculated  
\item[$^{\mathrm{b}}$] Equivalent width 
\item[$^{\mathrm{c}}$] Numbers in parenthesis are $1\sigma$ uncertainties, and
upper or lower limits correspond to $2\sigma$. 
\item[$^{\mathrm{d}}$] Partially contaminated by lines of other species (see text)
\end{list}
   \end{table}

   \begin{table}
      \caption{Line equivalent widths and fluxes of the SPIRE lines}
         \label{tab:spirefluxes}
\begin{center}
          \begin{tabular}{lccc}   
            \hline
            \noalign{\smallskip}
            Transition  & Velocities$^{\mathrm{a}}$    & $W_{\mathrm{eq}}$$^{\mathrm{b,c}}$  & Flux$^{\mathrm{c}}$ \\ 
            & (km\,s$^{-1}$)  & (km\,s$^{-1}$)     &  (Jy\,km\,s$^{-1}$) \\
            \noalign{\smallskip}
            \hline
            \noalign{\smallskip}
\ohp\ \, $1_0-0_1$                                        & $[ -100, 850]$   & $-146$ $( 18)$$^{\mathrm{d}}$   & $    293$ $(   36)$$^{\mathrm{d}}$   \\
                                       & $[-1200,-300]$   & $  65$ $( 18)$   & $   -130$ $(   36)$   \\
\ohp\ \, $1_2-0_1$                                        & $[ -100, 850]$   & $-145$ $( 15)$   & $    381$ $(   39)$   \\
                                       & $[-1200,-200]$   & $ 135$ $( 15)$$^{\mathrm{d}}$   & $   -354$ $(   39)$$^{\mathrm{d}}$   \\
\ohp\ \, $1_1-0_1$                                        & $[ -100, 850]$   & $ -79$ $( 27)$   & $    253$ $(   88)$   \\
                                       & $[ -730, -90]$   & $  67$ $( 27)$$^{\mathrm{d}}$   & $   -215$ $(   88)$$^{\mathrm{d}}$   \\
$\mathrm{oH_2O^+} \, 1_{11}\frac{1}{2}-0_{00}\frac{1}{2}$ & $[ -400, 400]$   & $ -65$ $( 10)$$^{\mathrm{d}}$   & $    268$ $(   42)$$^{\mathrm{d}}$   \\
            \noalign{\smallskip}
            \hline
         \end{tabular} 
\end{center}
\begin{list}{}{}
\item[$^{\mathrm{a}}$] Velocity range in which equivalent widths and fluxes
  are calculated  
\item[$^{\mathrm{b}}$] Equivalent width 
\item[$^{\mathrm{c}}$] Numbers in parenthesis are $1\sigma$ uncertainties
\item[$^{\mathrm{d}}$] Contaminated by lines of other species (see text)
\end{list}
   \end{table}


   \begin{figure*}
   \centering
   \includegraphics[width=14.0cm]{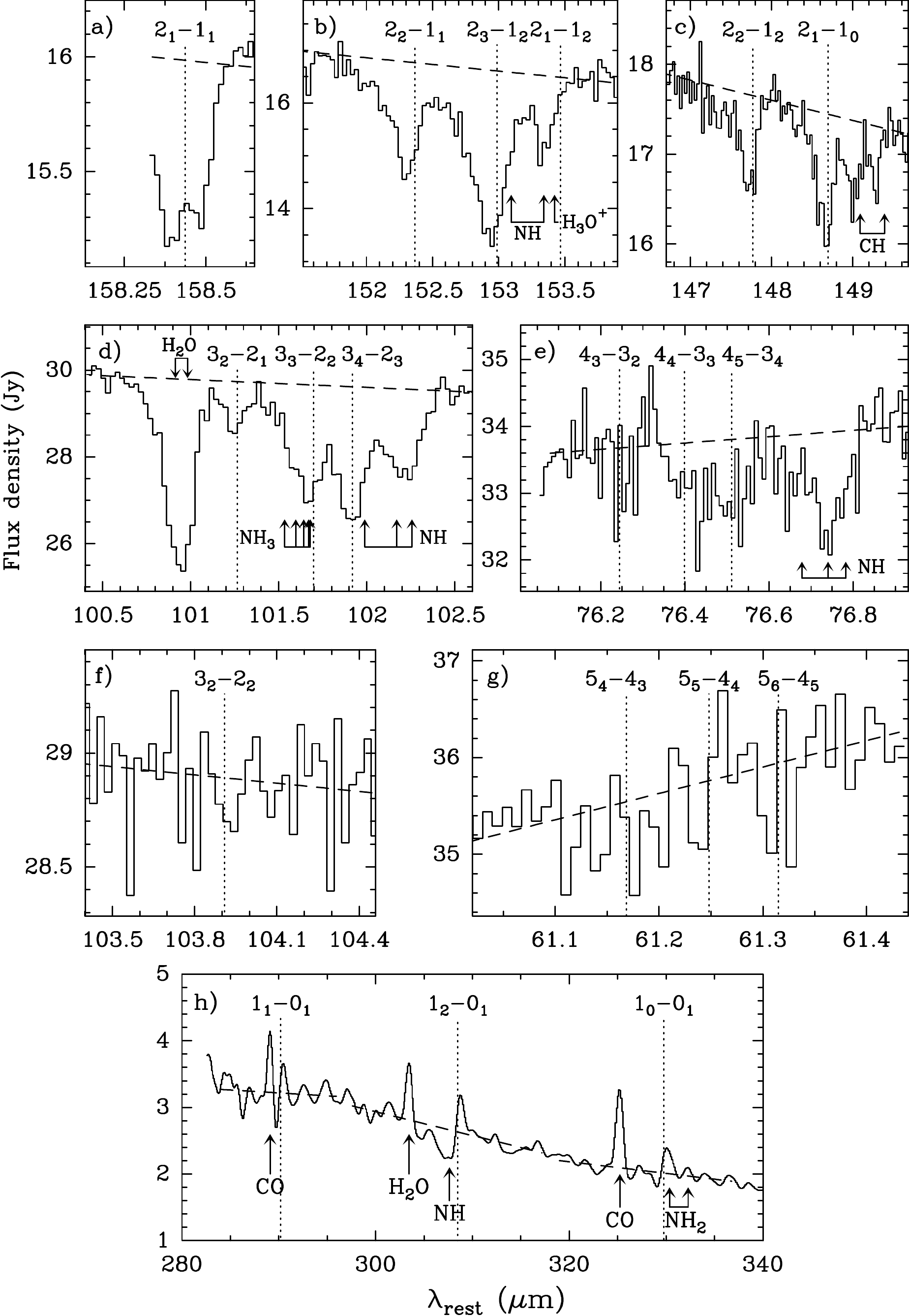}
   \caption{The observed spectra around the \ohp\ lines in Mrk~231, with
     the adopted baselines (dashed lines). The rest wavelengths, 
     marked by vertical dotted lines, are calculated
     with respect to the systemic redshift of $z=0.04218$. 
     Contributions to the spectra by NH (panels b, d, e and
     h), NH$_2$ (panel h), NH$_3$ (panel d), 
     CH (panel c), and CO (panel h) are also indicated. In panel
     a, the \ohp\ $2_1-1_1$ line is truncated at blueshifted velocities
     due to the proximity of the [C {\sc ii}] 158 $\mu$m line.
   }   
    \label{ohp-baselines}
    \end{figure*}

\subsection{The \ohp\ spectra}

The OH$^+$ observations, from both PACS (panels a-g) and SPIRE (panel h)
are displayed in 
Fig.~\ref{ohp-baselines}, together with the adopted baselines (dashed lines). 
In all cases, we have fitted polynomials of degree 1 around the (expected)
line position to minimize the uncertainties in the velocity extent of the line
wings. The most uncertain baseline corresponds to panel d, owing to the
proximity of the strong \hdo\ $2_{20}-1_{11}$ absorption line and, at shorter
wavelengths, the PACS 100 $\mu$m gap.

\subsubsection{Detections and potential contaminations}

Potential contamination by lines of other molecular species is indicated
in Fig.~\ref{ohp-baselines}. In panel a, the \ohp\ $2_1-1_1$ line is
truncated at blueshifted velocities due to the proximity of the [C {\sc ii}]
158 $\mu$m line; we nevertheless use the \ohp\ absorption at central
velocities to compare with model predictions in \S\ref{sec:analysis}.
In panel b, the \ohp\ $2_3-1_2$ line is closely blended with the NH $2_2-1_1$
line at $\approx153.1$ $\mu$m, as the NH $2_3-1_2$ line at $\approx153.35$
$\mu$m is clearly detected. Nevertheless, the overall profile is dominated by
\ohp\ as previously suggested based on {\it ISO} observations
\citep{gon08}. We also indicate in panel b the position of the very
high-lying \htop\ $12^--12^+$ transition ($E_{\mathrm{lower}}\approx1400$ K),
detected in Arp~220 (GA13) but absent in Mrk~231; the nearby
(intrinsically weak) \ohp\ $2_1-1_2$ line is also not detected. 
In panel c, we detect (blueshifted) absorption in the CH ground-state doublet
at 149 $\mu$m, but it is resolved from the neighboring \ohp\ $2_1-1_0$
line.

The redshift of Mrk~231 places the \ohp\ $3_J-2_{J'}$ lines 
($E_{\mathrm{lower}}\approx140$ K, $\lambda_{\mathrm{rest}}\sim102$\,$\mu$m) 
redward ($\lambda_{\mathrm{obs}}\sim106$\,$\mu$m) of the 100\,$\mu$m range
short wavelength leakage region of PACS, allowing 
its observation (Fig.~\ref{ohp-baselines}d)\footnote{In NGC~4418
  and Arp~220, these lines fell within the 100 $\mu$m gap and were thus
  not observed (GA13).}. Since NH has a
spectrum similar to \ohp, contamination is also found in the
\ohp\ $3_4-2_3$ line at $101.9$\,$\mu$m by the NH $3_2-2_1$ line at
$102.0$\,$\mu$m, as the blend of the NH 
$3_3-2_2$ and $3_4-2_3$ lines at $102.2$\,$\mu$m 
is well detected (panel d). Nevertheless, the
$3_3-2_2$ line of \ohp\ at $101.7$ $\mu$m is strongly detected with evidence
for blueshifted absorption, though contamination by the NH$_3$ $5_{K}-4_{K}$
lines (with $K=0,\ldots,4$) should be considered  
(see \S\ref{sec:analysis}). 
The strength of the intrinsically weak \ohp\ $3_2-2_1$ line at $101.3$
$\mu$m is doubtful owing to the proximity of the \hdo\ 101 $\mu$m line and the
uncertain continuum level. In panel e, the \ohp\ $4_5-3_4$ and $4_4-3_3$
lines are detected at systemic velocities, with hints of absorption in the
$4_3-3_2$ component; the \ohp\ $4_J-3_{J-1}$ lines are well separated from the
corresponding NH lines.  
In panels f and g, the spectra around the undetected \ohp\ $3_2-2_2$ and
$5_J-4_{J-1}$ lines are shown, and are used in \S\ref{sec:analysis} to check
for overall consistency with the models. Panel h shows the SPIRE spectrum
around the \ohp\ ground-state $1_J-0_1$ lines. The $1_1-0_1$ line is
contaminated on its blue side by the CO $9-8$ line at $289.12$ $\mu$m, the
$1_2-0_1$ line has potential contamination by the ground-state
NH $1_2-0_1$ line at $\approx307.6$ $\mu$m, and the emission component
of the $1_0-0_1$ line is most likely contaminated by 
NH$_2$ $2_{02}-1_{11}\,5/2-3/2$\footnote{First detected in the ISM by
    \cite{dis93}, NH$_2$ has a spectrum similar to H$_2$O$^+$ with the
    rotational levels of the asymmetric rotor split into 2 fine-structure
    levels due to the unpaired electronic spin of $1/2$.}.
The contributions by most of these species 
(NH, NH$_2$, NH$_3$, and CH) are included in our model for \ohp\ below
(\S\ref{oh+models}).

\subsubsection{Main characteristics of the \ohp\ lines}
\label{oh+charac}

   \begin{figure*}
   \centering
   \includegraphics[width=15cm]{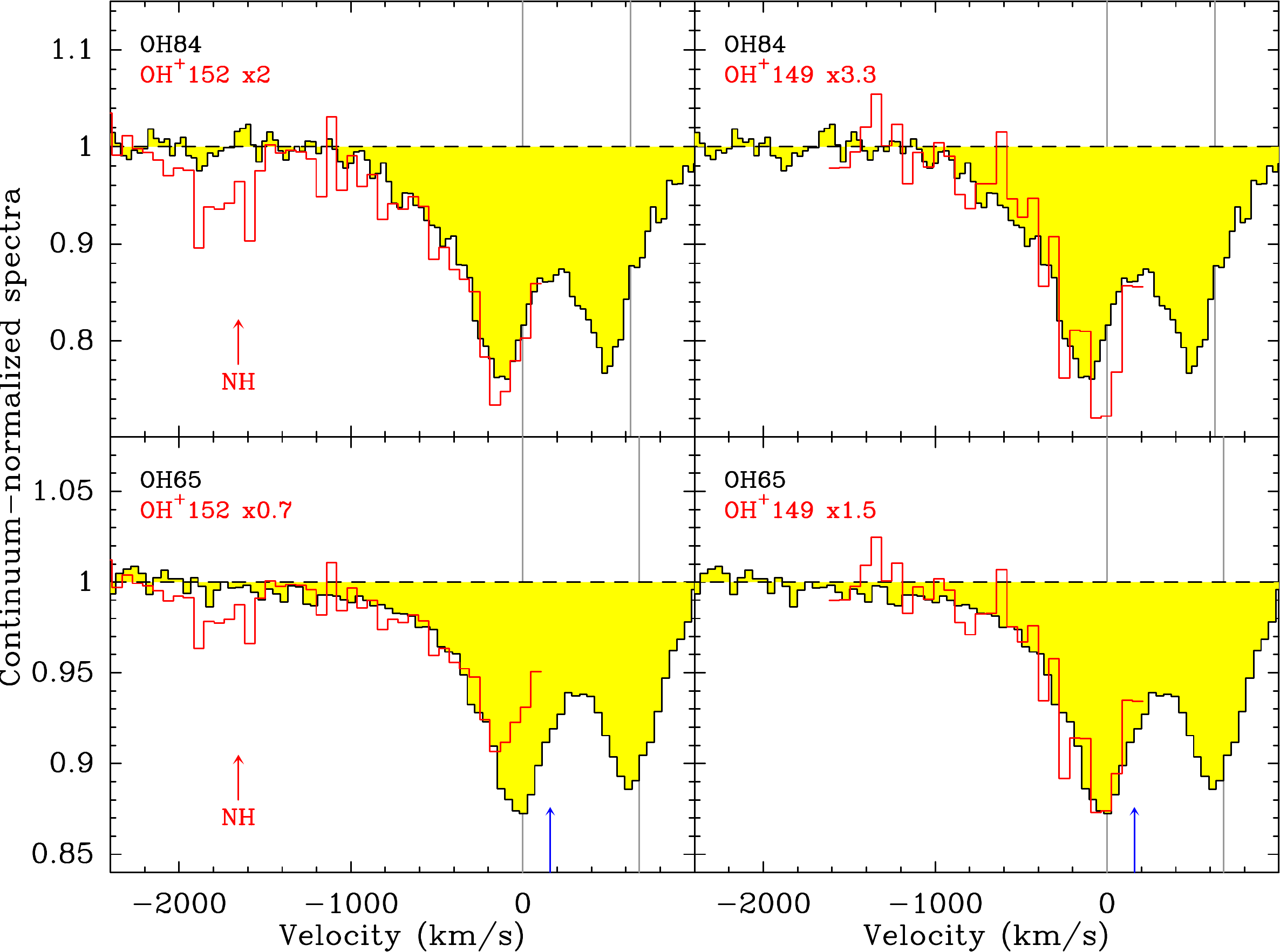}
   \caption{Comparison of the line shapes of \ohp\ $2_2-1_1$ at
     $152.4$ $\mu$m (left panels, in red) and \ohp\ $2_1-1_0$ at
     $148.7$ $\mu$m (right panels) with the 
     excited OH at 84 and 65 $\mu$m (yellow histograms). The two
     vertical gray lines in each panel indicate the positions of the 
     doublet components of the OH84 and OH65 transitions, and the
     blue arrows in the lower panels indicate potential contamination to the
     OH65 spectrum by the very excited H$_2$O $6_{25}-5_{14}$ line. The
     good match between the \ohp152 and OH84, and between the \ohp149 and OH65
     lines, at both central and blueshifted velocities
     indicates that the lines of both species are generated in the same
     regions.  
} 
    \label{comp_oh84_oh+152}
    \end{figure*}

The OH$^+$ spectra displayed in Fig.~\ref{ohp-baselines} show clear
indications of extremely high-velocity outflowing gas, together with
absorption near systemic velocities (as found for OH; GA14, GA17). In the
excited lines observed with PACS, the clearest outflow signatures are seen in
the relatively strong $2_2-1_1$ and 
$2_3-1_2$ lines, obtained with high signal-to-noise ratios (SNR;
Fig.~\ref{ohp-baselines}b). These lines have lower level energies of 
$E_{\mathrm{lower}}\approx50$~K (see Fig.~1 in GA13 for energy level diagrams
for the three O-bearing molecular ions). 
The relatively weak \ohp\ $2_2-1_2$ and $2_1-1_0$
lines at $\sim148$\,$\mu$m also show clear indication of blueshifted 
absorption (Fig.~\ref{ohp-baselines}c). 
The blueshifted line wings of the
uncontaminated $2_2-1_1$ $152.4$\,$\mu$m and $2_1-1_0$ $148.7$\,$\mu$m
transitions, showing absorption out to $\approx-1000$\,\kms\ from the line
center, closely match the line wings of the (also excited) OH84 and OH65
line profiles (Fig.~\ref{comp_oh84_oh+152}). At around systemic
velocities, the \ohp\ $2_2-1_1$ absorption also closely matches the OH84 line
profile, including the peak absorption blueshift of $\sim-120$ \kms, while the
optically thinner \ohp\ $2_1-1_0$ line better matches the OH65 profile.
This strongly suggests that both species are formed in the same gas
components.  The \ohp\ $2_3-1_2$ line profile also shows blueshifted
absorption reaching velocities of $-1200$\,\kms, merging with
the $2_2-1_1$ line (Fig.~\ref{ohp-baselines}b).

The $3_3-2_{2}$ and $3_4-2_{3}$ fine-structure components
(Fig.~\ref{ohp-baselines}d), with $E_{\mathrm{lower}}\approx140$~K, also show
indications of blue absorption, as the potential contamination by \nht\  
to the blueshifted wing of the $3_3-2_{2}$ line is expected to be weak
(\S\ref{sec:analysis}).  No clear blueshifted absorption in the \ohp\ profiles
is seen in the $4_J-3_{J'}$ transitions (Fig.~\ref{ohp-baselines}e,
$E_{\mathrm{lower}}\approx280$\,K).

The ground-state \ohp\ $1_J-0_1$ lines detected with SPIRE also show evidence
for outflowing gas (Fig.~\ref{ohp-baselines}h), with the emission above the
continuum redshifted in the 
three lines relative to the expected line centers, and also with clear
indications of blueshifted absorption. The latter is seen in the
$1_2-0_1$ line profile although NH probably also contributes to the
broad absorption feature (\S\ref{sec:analysis}), and 
in the $1_1-0_1$ line profile in spite of the proximity of the CO $9-8$ line.
Some hints of absorption are also observed in the weakest transition,
$1_0-0_1$.

As noted above, the strongest low-lying $2_J-1_{J'}$ lines in
Fig.~\ref{ohp-baselines}b-c show the peak absorption blueshifted by $\sim120$
\kms, while the higher-lying lines peak at central velocities.  This behavior
resembles that found in the OH lines of Mrk~231 (GA14), where the peak
absorption of the OH84 transition is also blueshifted but the higher-lying
lines (e.g., OH65 and OH71) peak closer to the systemic velocities
(see also Fig.~\ref{comp_oh84_oh+152}). In addition, no excited \ohp\ lines
show redshifted emission above the continuum.  This behavior is also similar
to that seen in the OH line profiles.  Only the ground-state OH119 and OH79
doublet profiles show prominent redshifted emission (GA14, GA17), just as only
the ground-state lines of \ohp\ show redshifted emission features
\citep[Fig.~\ref{ohp-baselines}h;][]{wer10}. Finally, comparison of the
\ohp\ PACS spectra in Mrk~231 and Arp~220 indicates that the absorption
strength (relative to the continuum) at systemic velocities is similar in both
sources (Fischer et al., in prep.), the important difference being the
high-velocity blueshifted absorption in Mrk~231 that is absent in Arp~220.

Because of the non-Gaussian shapes of the \ohp\ lines, we  
measured the equivalent widths and fluxes of the PACS lines in 
Table~\ref{tab:fluxes} by numerically integrating the line shapes over two
velocity ranges, around systemic velocities and, when 
detected, in the blueshifted wings. 
Based on the change in the slope of the 
  \ohp\ PACS spectra (see e.g. Fig.~\ref{comp_oh84_oh+152}), 
we adopted blueshifted and systemic velocity ranges of $[-1000,-300]$ and 
$[-300,200]$\,km\,s$^{-1}$, respectively; the latter upper limit is set 
to avoid strong contamination by neighboring lines on the red side.
The $1\sigma$ uncertainties of the equivalent widths were evaluated 
from the $\sigma_{\mathrm{rms}}^{\mathrm{norm}}$ noise of the continuum-normalized
PACS spectra as $\sigma_{\mathrm{rms}}^{\mathrm{norm}}\times\Delta V/n^{1/2}$, where
$\Delta V$ and $n$ are the velocity coverage and number of channels over which
$W_{\mathrm{eq}}$ is measured \citep{gon15}. Line detection is established at
$\geq3\sigma$ level.
The equivalent widths and fluxes of the \ohp\ ground-state lines
(Table~\ref{tab:spirefluxes}) as measured with SPIRE are also
integrated over two velocity ranges, corresponding to the emission and
absorption features. We adopted a common velocity range
  ($[-100,850]$\,km\,s$^{-1}$) for the emission features, enabling comparison
  of the strengths of the \ohp\ $1_{J}-0_1$ emission lines. Uncertainties for
  the equivalent widths of the ground-state \ohp\ lines are dominated by the
  remaining ripples of the apodized spectrum, rather than by random noise. A
  typical semi-cycle of the  
  ripples has a width similar to that of the observed lines, so that we
  conservatively adopt the area of one semi-cycle (averaged over several
  cycles measured close to the line)
  as the $1\sigma$ uncertainty of the equivalent width, and adopt a detection
  criterion of $\geq2\sigma$.

We caution that with the current
spectral resolution of the apodized spectrum ($\sim650$\,km\,s$^{-1}$) the
\ohp\ ground-state emission features will be contaminated by the
absorption and vice versa, and that the ripples in the spectrum 
(Fig.~\ref{ohp-baselines}h) make our flux measurements
somewhat uncertain (Table~\ref{tab:spirefluxes}). 
Nevertheless, optically thin emission would imply relative
fluxes for the $1_0-0_1$, $1_1-0_1$, and $1_2-0_1$ lines of $1:3:5$,
which are the degeneracies of the upper levels\footnote{This strictly
    applies to collisional excitation and, most probably, to the emission
    associated with formation pumping; if the \ohp\     
    scatters continuum radiation, the contrast between the $1_0-0_1$
    component and the other two would be even higher due to the decreasing
    continuum strength with increasing wavelength.}. 
The \ohp\ $1_0-0_1$ and
$1_1-0_1$ emission features show, however, similar fluxes, which are a
factor $\sim1.4$ weaker than the $1_2-0_1$ emission 
(Table~\ref{tab:spirefluxes}). This suggests both strong
contamination of the $1_0-0_1$ line by NH$_2$\footnote{The emission 
feature at $907.4$ GHz in the \ohp\ $1_0-0_1$ spectrum of Arp~220 shown by 
\cite{ran11} is most probably dominated by NH$_2$ $2_{02}-1_{11}\,5/2-3/2$, as 
the $3/2-1/2$ component at $902.2$\,GHz is clearly seen in their spectrum.}, 
and significant optical depth effects in the $1_J-0_1$ lines.

\subsection{The \hdop\ spectra}
\label{sec:h2opspec}

\begin{figure*}
   \centering
   \includegraphics[width=15.0cm]{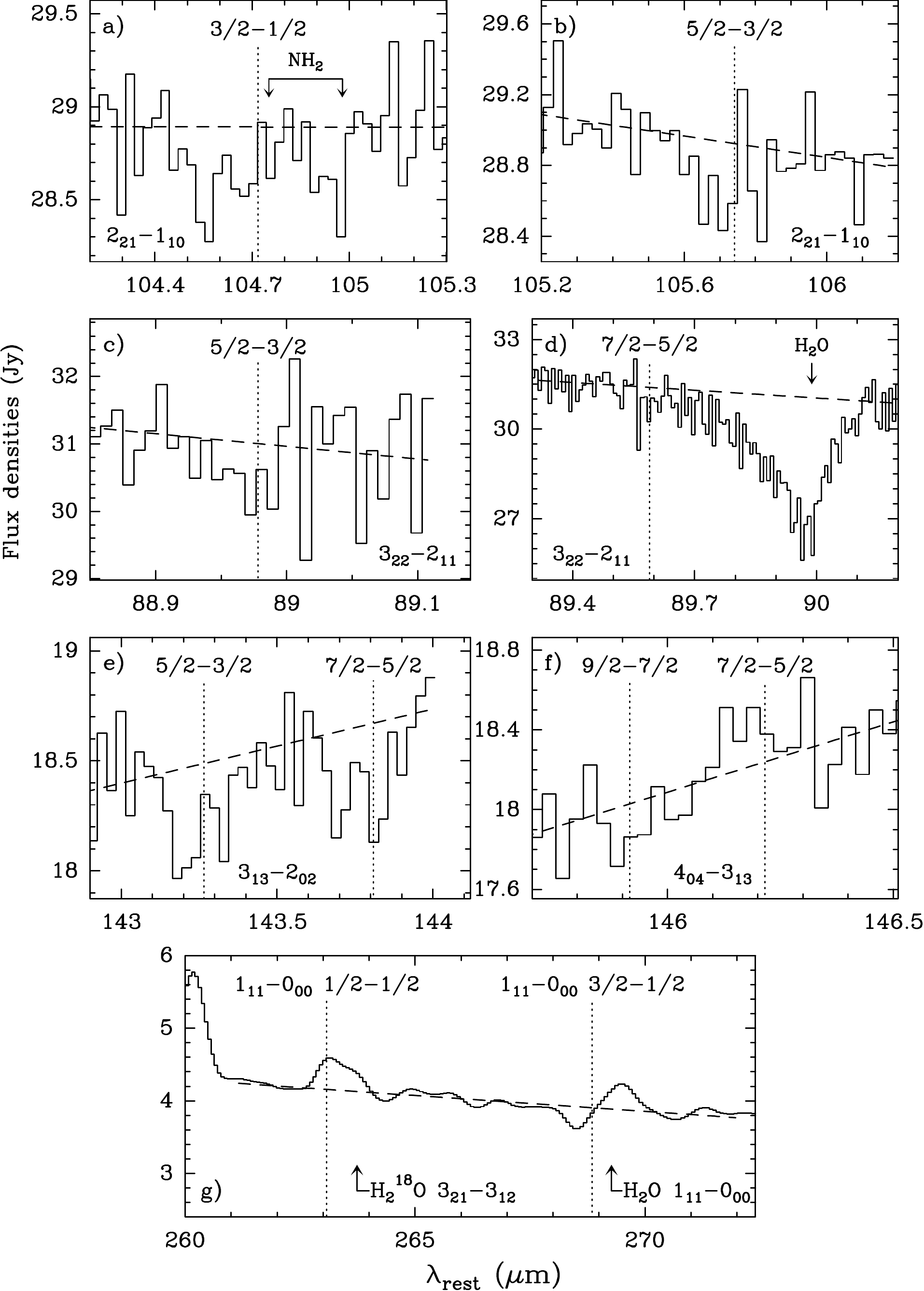}
   \caption{The observed spectra around the uncontaminated and presumably
     strongest (as seen in Arp~220, GA13) \hdop\ lines in Mrk~231, with
     the adopted baselines (dashed lines). The rotational transitions are
     indicated at the bottom of each panel, and the fine-structure
     components are indicated above the spectra. The rest wavelengths are
     calculated with respect to the systemic redshift of $z=0.04218$. 
      The absorption shown in panel a is probably contaminated by NH$_2$
       $3_{22}-2_{11}\,5/2-3/2$ and $7/2-5/2$. The 
     ground-state \hdop\ $1_{11}-0_{00}$ $1/2-1/2$ line in panel g has a
     redshifted shoulder attributable to H$_2^{18}$O, and the $3/2-1/2$
     fine-structure component is strongly blended with the para-H$_2$O
     $1_{11}-0_{00}$ ground-state line. 
   }   
    \label{h2op-baselines}
    \end{figure*}

In GA13, the strongest, lowest-lying, and uncontaminated lines of 
\hdop\ in the far-IR were identified and analyzed in Arp~220 and NGC~4418. We
show in Fig.~\ref{h2op-baselines}a-f\footnote{In Mrk~231, the ground-state
para-\hdop\ $2_{12}-1_{01}$ lines at rest wavelengths of $182.9$ and
$184.3$\,$\mu$m were not covered by our full PACS scan of Mrk~231 as 
  they lie in the $\gtrsim190$ $\mu$m leakage region.} the PACS spectra of
Mrk~231 around the same lines (the spectrum in panel a is
 contaminated by NH$_2$). 
Evidence for absorption in Mrk~231 is
found in three \hdop\ transitions: the $2_{21}-1_{10} \,\, 5/2-3/2$
para-\hdop\ line (Fig.~\ref{h2op-baselines}b), and the 
$3_{13}-2_{02} \,\,5/2-3/2$ and $7/2-5/2$ ortho lines (panel e). There may
also be some hints of absorption in the $3_{22}-2_{11} \,\, 5/2-3/2$ line
(panel c), but the $7/2-5/2$ fine-structure component, which is expected to be
stronger, is not detected (panel d). 

The detected \hdop\ lines peak close to
systemic velocities with no evidence for blueshifted absorption wings. 
In comparison with Arp~220, the $2_{21}-1_{10} \,\, 5/2-3/2$ and 
$3_{13}-2_{02} \,\,5/2-3/2$ lines have similar absorption strengths (relative
to the continuum) in both sources, but the $3_{13}-2_{02} \,\,7/2-5/2$ line
is stronger in Arp~220.

The SPIRE spectrum shows strong emission in the ground-state \hdop\ 
$1_{11}-0_{00} \,\, 1/2-1/2$ line (Fig.~\ref{h2op-baselines}g), which is
surprising given the weakness of the absorption lines seen with PACS. While
the emission feature has a peak strength similar to that of the
\ohp\ $1_1-0_1$ line (Fig.~\ref{ohp-baselines}h), it shows no measurable
absorption at blueshifted velocities -- in contrast with the
\ohp\ ground-state lines even though the continuum is stronger 
at the wavelength of the \hdop\ line. On the other hand, the 
\hdop\ $1_{11}-0_{00} \,\, 1/2-1/2$ spectral feature shows a prominent
redshifted shoulder, which is attributable to the H$_2^{18}$O $3_{21}-3_{12}$ 
line. This identification relies on the facts that the corresponding
$3_{21}-3_{12}$ line of the main isotopologue H$_2^{16}$O is strong in Mrk~231
\citep{gon10}, and that $^{18}$O is extremely abundant in this source as
inferred from PACS observations and modeling of OH and $^{18}$OH 
\citep[][GA14]{fis10}. 
On the other hand, the feature identified with \hdop\ cannot be a blueshifted
feature of the H$_2^{18}$O line, because the corresponding H$_2^{16}$O line
shows no blueshifted shoulder. Unfortunately, the \hdop\ 
$1_{11}-0_{00} \,\, 3/2-1/2$ fine-structure component
(Fig.~\ref{h2op-baselines}g) is blended with the
ground-state $1_{11}-0_{00}$ line of H$_2^{16}$O, and cannot confirm the
interpretation of the $1/2-1/2$ line. Our searches of the CDMS and JPL
catalogs, however, do not reveal plausible alternatives to the quoted
\hdop\ and H$_2^{18}$O lines to explain the spectral emission at 
$263-264$\,$\mu$m.

Tables~\ref{tab:fluxes} and \ref{tab:spirefluxes} list the equivalent
  widths and fluxes of the \hdop\ PACS and SPIRE lines,
  respectively. Uncertainties were calculated in the same way as for the
  \ohp\ lines (\S\ref{oh+charac}). The flux of the ground-state 
  $1_{11}-0_{00} \,\, 1/2-1/2$ line was integrated up to $400$\,\kms\ to avoid
  strong contamination by the blended H$_2^{18}$O line.

\subsection{The \htop\ spectra}
\label{sec:h3opspec}

\begin{figure*}
   \centering
   \includegraphics[width=15.0cm]{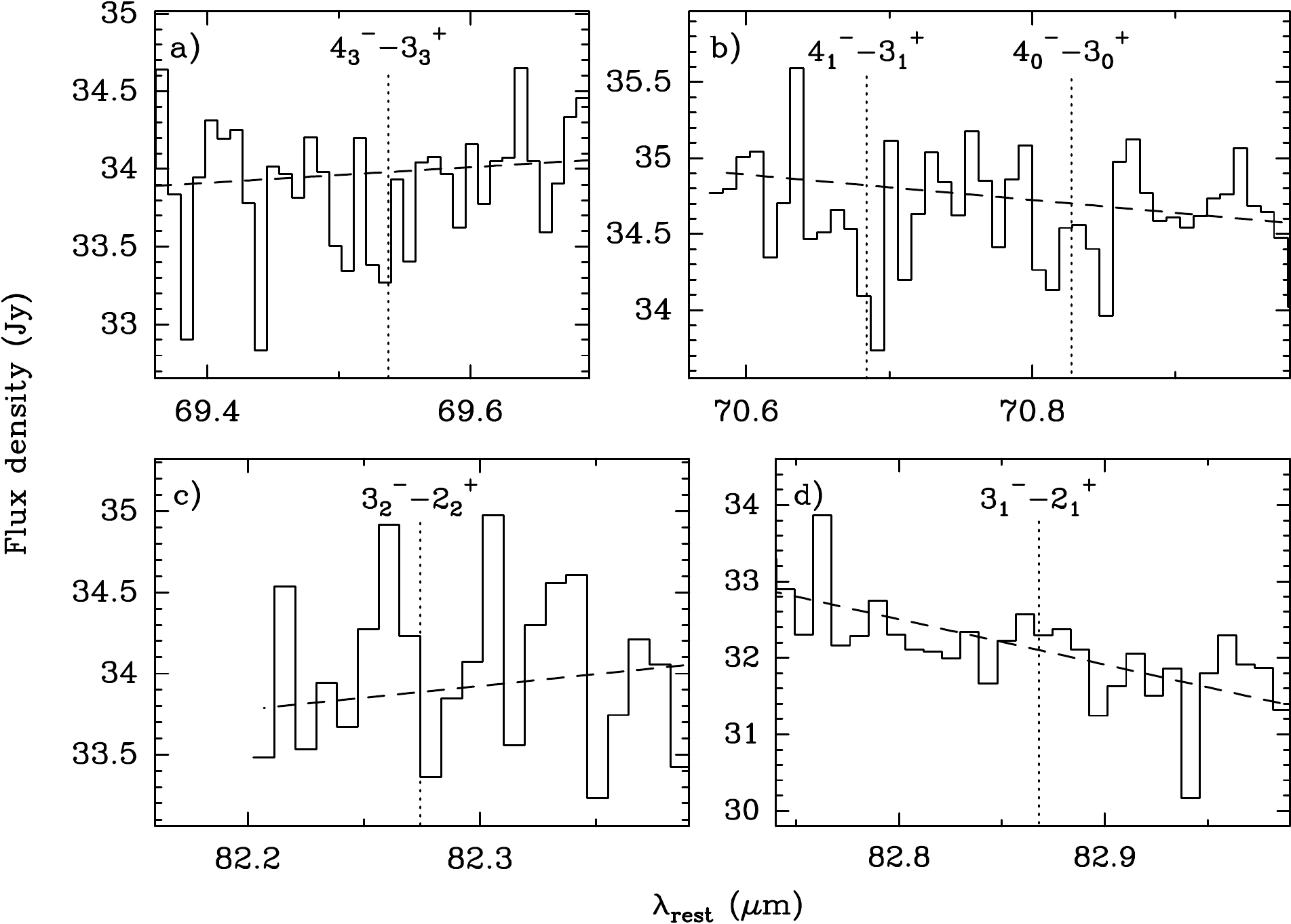}
   \caption{The observed spectra around the uncontaminated and presumably
     strongest (as seen in NGC~4418 and Arp~220, GA13) \htop\ lines in
     Mrk~231, with the adopted baselines (dashed lines). The rest wavelengths
     are calculated with respect to the systemic redshift of
     $z=0.04218$. There are only some hints of absorption in the lines shown
     in panels a and b, and no hints of absorption in the
     pure-inversion metastable lines (not shown). Likewise, the ground-state
     ortho line $0_0-1_0$ line at $304.45$ $\mu$m is not detected (not shown).
   }   
    \label{h3op-baselines}
    \end{figure*}

There is very little evidence for \htop\ in the PACS and SPIRE
spectra of Mrk~231. A search for the transitions that were detected in
Arp~220 and/or NGC~4418 (GA13) resulted mostly in negative results. 
The lower signal-to-noise ratio in some ranges of the Mrk~231 spectrum in
comparison with that of Arp~220, cannot on its own account for this
difference.  
In contrast to Arp~220, no pure-inversion metastable lines are detected in 
Mrk~231. We show in Fig.~\ref{h3op-baselines} the PACS spectrum of Mrk~231
around the \htop\ lines that are expected to be strongest as seen in the
PACS spectrum of Arp~220. The only transition that shows some
hints of absorption is the ortho-\htop\ $4_3^--3_3^+$. We use these spectra in
\S\ref{sec:analysis} to establish an upper limit for the
\ohp/\htop\ ratio. The ortho-\htop\ ground-state $0_0^--1_0^+$ line at
$304.45$ $\mu$m is also absent in Mrk~231 (not shown). 
Upper limits for the equivalent widths and fluxes of the \htop\ PACS
  lines ($2\sigma$, calculated in the same way as for the \ohp\ lines) are
  listed in Table~\ref{tab:fluxes}.  

\section{Analysis}
\label{sec:analysis}

\subsection{Radiative transfer models}

   \begin{figure*}
   \centering
   \includegraphics[width=15.0cm]{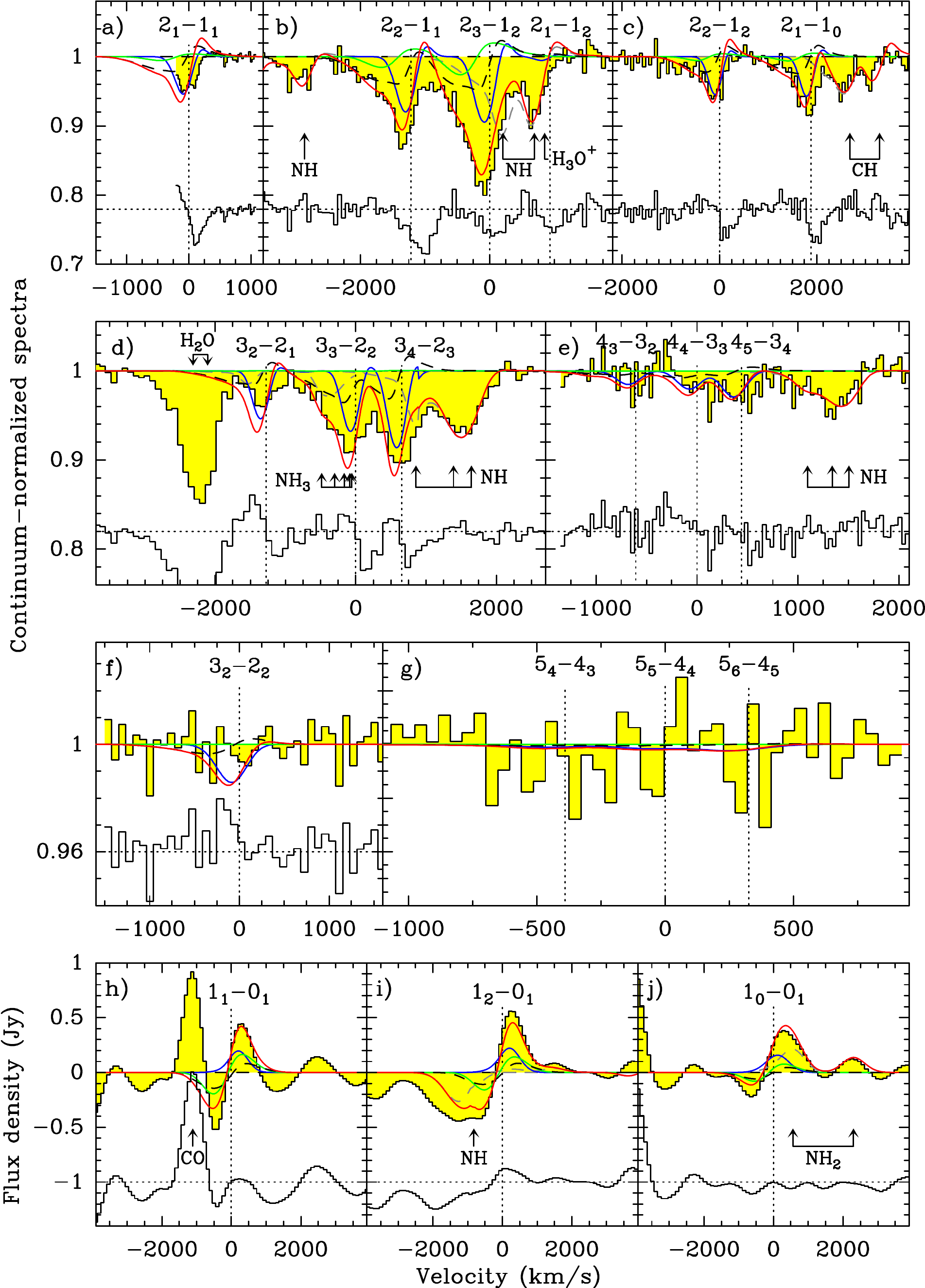}
   \caption{Model fit for the \ohp\ lines in Mrk~231. The blue, dashed-black, 
     and green curves show the contributions by the quiescent component (QC),
     the high-excitation outflow component (HVC), and the low-excitation
     component (LEC), respectively. The model includes the estimated
     contributions by NH (panels b, d, e, i), NH$_2$ (panel j), NH$_3$
     (panel d), and CH (panel c), shown as gray-dashed curves. The red curve
     indicates the summed profiles from all three components, and the
       black histograms at the bottom of each panel (except g) show the
       residuals (data$-$model) vertically shifted by the value indicated with
       the horizontal-dotted lines. The physical
     parameters used for the QC and HVC models are the 
     same as in the model for OH in GA17. In panel a, the
     \ohp\ $2_1-1_1$ line is truncated at blueshifted velocities due to the
     proximity of the [C {\sc ii}] 158 $\mu$m line.
   } 
    \label{fitoh+}
    \end{figure*}

   \begin{figure}
   \centering
   \includegraphics[width=8.5cm]{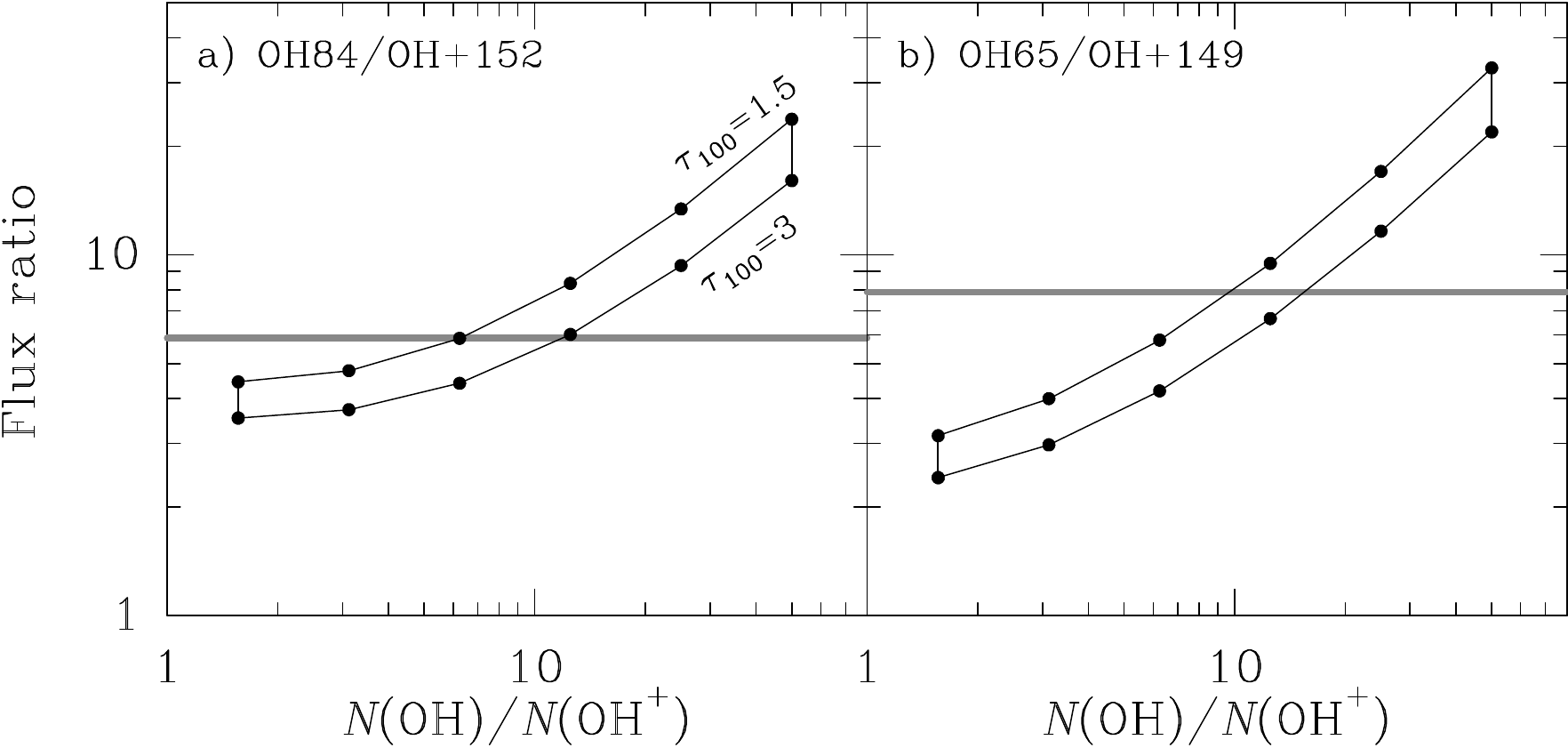}
   \caption{a) The OH84/OH+152 and b) the OH65/OH+149 line flux ratios (both
     in erg\,s$^{-1}$\,cm$^{-2}$),
     integrated between $-1000$ and $-300$\,\kms, as a function of the
     OH/\ohp\ column density ratio. Model results for the high-velocity
     component (HVC) are shown for optical depths
     of the continuum source at 100\,$\mu$m of $\tau_{100}=1.5$ and $3$. The
     horizontal lines show the values measured in Mrk~231, indicating
     $N(\mathrm{OH})/N(\mathrm{OH^+})\approx10$ in the HVC. Therefore,
     $X(\mathrm{OH^+})>10^{-7}$  for $X(\mathrm{OH})>10^{-6}$ in the
     excited, outflowing component.
   } 
    \label{oh_ohplus}
    \end{figure}

   \begin{figure*}
   \centering
   \includegraphics[width=15.0cm]{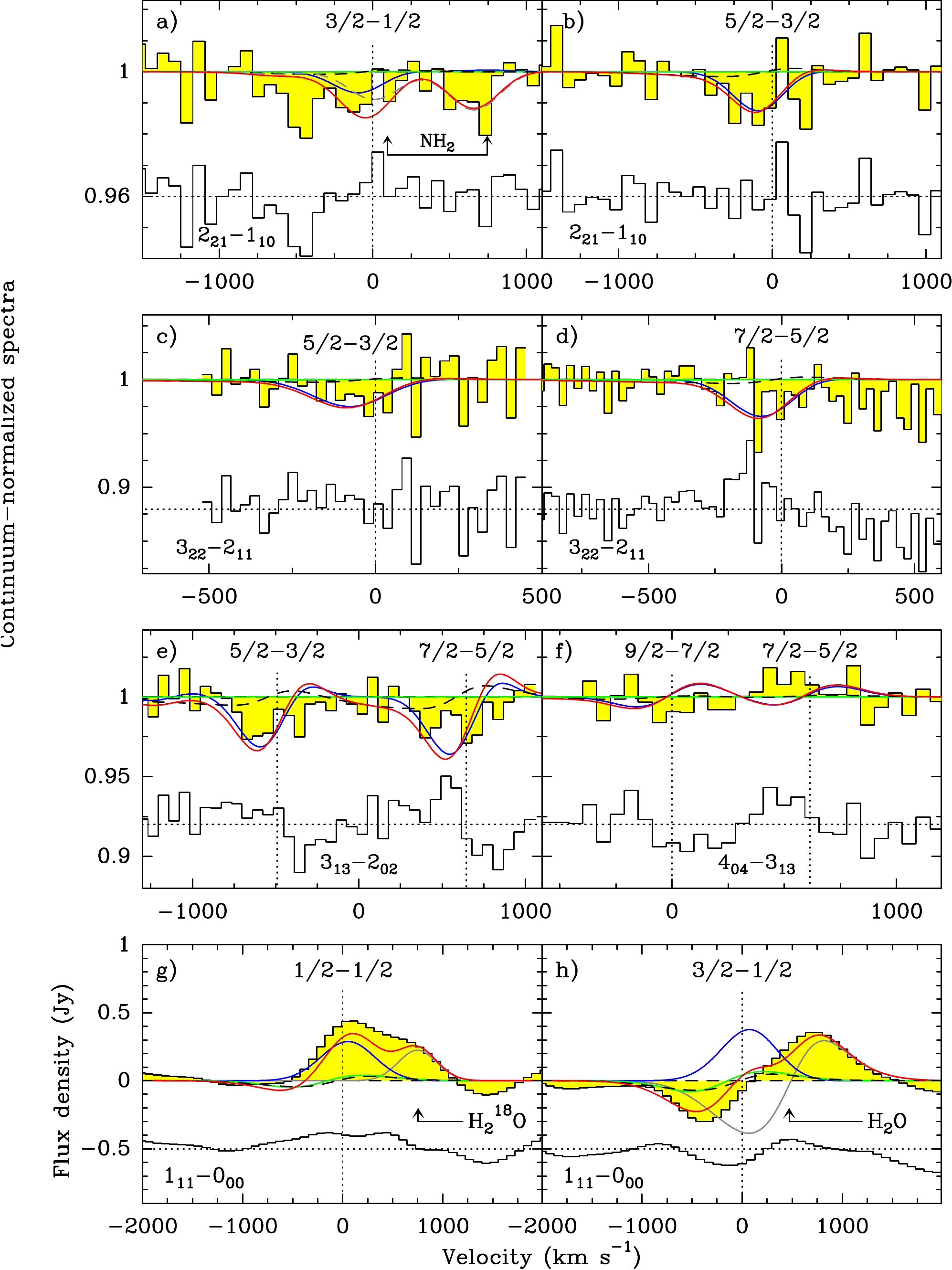}
   \caption{A model for the \hdop\ lines in Mrk~231, which 
     establishes a lower limit for the \ohp/\hdop\ abundance ratio in both the
     QC (blue lines) and the HVC (dashed-black lines). 
     The LEC is also included
     in green, but has a negligible effect on the PACS lines. 
     The red curve
     indicates the summed profiles from all three components, and the
       black histograms at the bottom of each panel show the
       residuals (data$-$model) vertically shifted by the value indicated with
       the horizontal-dotted lines. We obtain
     \ohp/\hdop$\gtrsim4$ for the QC and \ohp/\hdop$\gtrsim10$ for the HVC
     (mostly based on the detected lines in panel e). The physical
     parameters used for the QC and HVC models are the 
     same as in the model for OH in GA17. Panel a includes a model for
       NH$_2$ (the same as used in Fig.~\ref{fitoh+}j), panel g includes
       H$_2^{18}$O, and panel h includes a model  
     for H$_2$O in the outflow (in gray).
   } 
    \label{fith2o+}
    \end{figure*}

   \begin{figure*}
   \centering
   \includegraphics[width=15.0cm]{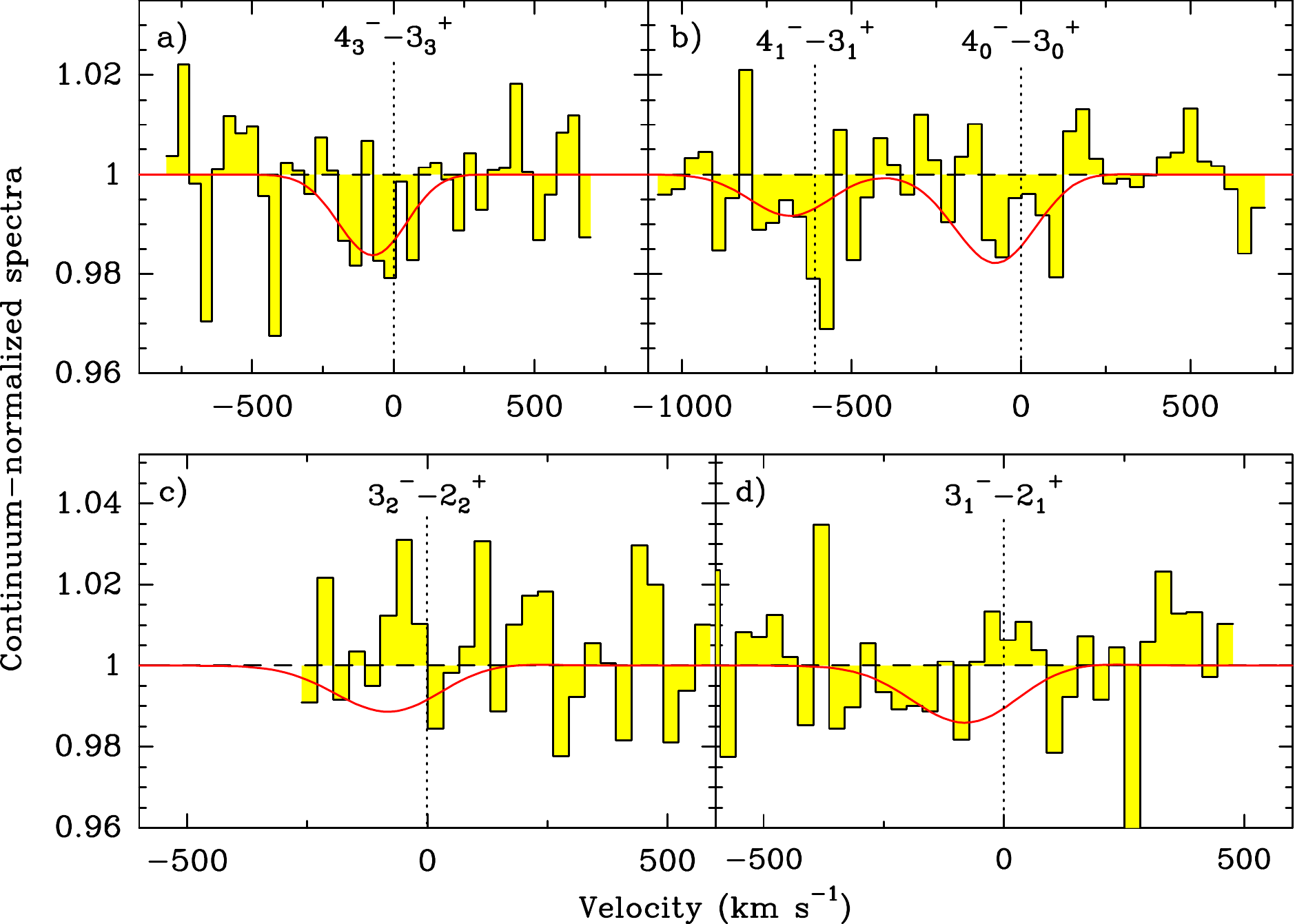}
   \caption{A model for the \htop\ lines in Mrk~231, which 
     establishes a lower limit for the \ohp/\htop\ abundance ratio in the
     QC (no significant limits are obtained for the outflowing components). We
     obtain \ohp/\htop$\gtrsim4$ for the QC. The physical
     parameters used for the QC model are the 
     same as in the model for OH in GA17.
   } 
    \label{fith3o+}
    \end{figure*}

GA17 modeled the OH line profiles in Mrk~231 with three components,
(i.e., a simpler scheme than the four-component model of GA14). 
The ``CORE'' component, which hereafter will be called the quiescent
component (QC), physically represents the $\sim100$ pc-scale torus
around the AGN detected in OH at centimeter wavelengths \citep{klo03},
and is intended to account for the line absorption 
in the nuclear region of the galaxy close to systemic velocities. 
The line wings and P-Cygni profiles observed in OH require two additional
``ENVELOPE'' components to reproduce spectral features arising in the
outflowing molecular gas. One of the outflowing components is compact, as it
accounts for the absorption in the excited OH84 and OH65
doublets. It has maximum velocities of $\gtrsim1000$ km s$^{-1}$, and will be
referred to as the high-velocity component (HVC). 
The other outflow component has lower 
maximum velocities and is primarily detected in the ground-state
OH119 and OH79 doublets; it will be referred to as the low-excitation
component (LEC). 
In order to account for most of the absorption and emission in OH119 and OH79,
the LEC must be more extended than the QC and HVC components, but it
contributes little to the excited OH84 and OH65 doublets. The spectra of the
O-bearing molecular ions reported in the present work are modeled and
interpreted in terms of this physically and spectrally motivated 3-component
model. By using the same physical parameters for these components
as in GA17, our only free parameters here are the abundances of \ohp, \hdop,
and \htop\ relative to H nuclei 
(regardless of whether the hydrogen is in atomic or molecular form), 
i.e., $X(\mathrm{OH^+})$, $X(\mathrm{H_2O^+})$, and $X(\mathrm{H_3O^+})$
in each component (QC, HVC, and LEC). Because the physical model
(continuum optical depths, dust temperatures, sizes, densities, linewidths,
velocity gradients, molecular distribution) takes exactly the same values for
all modeled species, the abundance ratios are less dependent on intrinsic
model uncertainties. 

The molecules and dust are mixed in the CORE components, and hence the
absolute abundances for the QC are inferred from the molecular
column density and the continuum optical depth at 100 $\mu$m,
$\tau_{100}$, by assuming a standard gas-to-dust ratio of 100 by mass
\citep[see also][]{gon12,fal15,sto17}. For 
the outflowing components, however, the optical depth of the mixed dust is not
well constrained (the absorption is primarily produced by the continuum source
behind the approaching gas), and an abundance
$X(\mathrm{OH})=2.5\times10^{-6}$ was adopted in GA14 and GA17.

For a given species, the column densities associated with the QC, HVC, and LEC
components are derived from a combination of kinematic and excitation
characteristics. The spectral features (in absorption or emission) detected at
central velocities are associated with the QC. The line wings associated with
{\it excited} lines are associated with the HVC, with specific predictions for
the absorption and emission in the ground-state lines. Therefore, the line
wings associated with the ground-state lines, which similarly to OH arise
mainly in a more extended and less excited component, are reproduced once the
LEC is included in the full model. 

\subsubsection{Models for \ohp}
\label{oh+models}

Figure~\ref{fitoh+}a-g presents our best model fit to the \ohp\ line profiles
observed by PACS and Table~\ref{tab:col} lists the inferred \ohp\ column
densities and abundances. Besides the pumping generated by the strong far-IR
radiation field, we also include collisional excitation with H atoms
\citep{liq16} and with electrons \citep{tak13} assuming a fractional abundance
$X(\mathrm{e^{-}})=10^{-3}$ (see \S\ref{sec:chemqc}); collisional excitation
with H is found to dominate. In our model fitting, we also include predictions
for NH, NH$_2$, NH$_3$, and  CH.  We find
similar NH and \ohp\ column densities for the QC, and we adopt the same
column for NH as for \ohp\ in the outflowing components as well.  Our model
for NH$_2$ and NH$_3$ only includes the prediction for the QC, based on
  the fits to other unblended NH$_2$ and NH$_3$ lines detected across the 
  PACS spectrum (Fischer et al.\ in prep).  

The PACS data mainly constrain the column densities of the QC and HVC
components, where \ohp\ is excited by the far-IR field. 
Model results for the HVC are shown in Fig.~\ref{oh_ohplus}, which compares
the predicted OH84/\ohp152 and OH65/\ohp149 flux ratios in the
blueshifted wings (the line observations that are compared in
Fig.~\ref{comp_oh84_oh+152}) with the measured 
values, indicating OH/\ohp\ column density ratios of $\sim10$ in the excited
outflow component. For the QC, the OH/\ohp\ ratio is $\sim20$
(Table~\ref{tab:col}). These column density ratios indicate very high
\ohp\ abundances, $\sim10^{-7}$ in the QC and even higher in the HVC for
$X(\mathrm{OH})>10^{-6}$ \citep{sto17}. 

Our model for the QC, which includes an expansion velocity of 100
\kms\ (GA17), reproduces the blueshift observed in the \ohp\ $2_J-1_{J'}$
lines but also predicts a blueshift in higher-lying lines that is not
observed.  The quality of the
fit is nevertheless reasonable for most lines at central and blueshifted
velocities, though our spherically symmetric models predict significant
redshifted reemission above the continuum in some $2_J-1_{J'}$, which is not
detected (see the residuals in Fig.~\ref{fitoh+}).  In addition, an
observed ``bridge'' of absorption joins the  
$2_3-1_{2}$ and the $2_2-1_{1}$ lines (panel b), and the $3_3-2_{2}$ and the
$3_2-2_{1}$ lines (panel d) as well, that cannot be reproduced.
These deviations of the spectral fits from the data may
indicate a departure from spherical symmetry, with the outflow mainly focussed
along the line of sight (GA14), or a redshifted absorption component that
could represent low-velocity inflowing gas.

Figs.~\ref{fitoh+}h--j compare the profiles of the ground-state \ohp\ lines
observed with SPIRE with the predictions of our composite model. We have not
attempted to correct the spectrum for the residual ripples in the 
FTS apodized spectrum, and thus our results are only approximate. 
Two main conclusions can nevertheless be drawn from this comparison.
First, the HVC and the QC together cannot account for either the absorption or
the emission features observed in the \ohp\ $1_J-0_{1}$ P-Cygni profiles, so
that we require the inclusion in the model of the LEC with a high \ohp\ column
of $\sim10^{16}$ cm$^{-2}$. This is a factor of only $\sim5$ lower 
than the column derived for OH in the same LEC, and thus a very high
\ohp\ abundance ($>2\times10^{-7}$) is favored. Second, the observed
\ohp\ redshifted emission features, including in particular that of the
uncontaminated $1_2-0_{1}$ component (Fig.~\ref{fitoh+}i), require high
volume densities (i.e. collisional excitation) at least for the
QC. Otherwise, the model for the QC would 
predict the \ohp\ ground-state lines in absorption close to systemic
velocities, which is not observed, and the emission features would be 
underpredicted. In the model shown in Fig.~\ref{fitoh+}, the density 
decreases from $\sim10^6$ cm$^{-3}$ in the innermost regions of the QC to
$\sim1.5\times10^5$ cm$^{-3}$ in the external parts. 


Even with high \ohp\ column densities in both outflowing components,
  Figs.~\ref{fitoh+}h--j show that the absorption features of the ground-state
  \ohp\ lines, and in particular the $1_0-0_{1}$ one, are underestimated by
  the model. The three lines are 
  optically thick in the three model components, so that an increase in
  column density would hardly improve the fit. The most probable reason for
  this discrepancy is the model underestimation of the
  submillimeter continuum by a factor of $\approx2$ at $300$\,$\mu$m;
a moderate increase in \tdust\ or in the continuum optical depth at
submillimeter wavelengths would produce deeper absorption in the $1_J-0_{1}$
lines.  
On the other hand, we also note that the high densities inferred from the
ground-state emission features, which are volume tracers as they probe deep
into the buried nuclear regions, do not necessarily 
apply to the gas responsible for the absorption features in excited lines, 
which are surface tracers; absorption and emission features are indeed
produced in different regions of the QC (see discussion in 
\S\ref{sec:chemqc}).


   \begin{table*}
      \caption{Physical parameters for the three model components of
        Mrk~231}  
         \label{tab:col}
          \begin{tabular}{lcccccccccc}   
            \hline
            \noalign{\smallskip}
Component & $T_{\mathrm{dust}}$$^{\mathrm{a}}$ & $\tau_{100}$$^{\mathrm{b}}$ & $N$(OH)$^{\mathrm{c}}$  & $N$(\ohp)  & $N$(\hdop)  & $N$(\htop)$^{\mathrm{d}}$  & $X$(OH)$^{\mathrm{e}}$  & $X$(\ohp)  & $X$(\hdop)  & $X$(\htop) \\  
          & (K) & & ($10^{16}$ cm$^{-2}$) & ($10^{16}$ cm$^{-2}$) & ($10^{16}$
cm$^{-2}$) & ($10^{16}$ cm$^{-2}$) &   & & &        \\
            \noalign{\smallskip}
            \hline
            \noalign{\smallskip}
QC  & 90 & 3.0 & $675$  & $34$  & $4.2-8.4$  & $\lesssim8.4$ &
$1.7\times10^{-6}$ & $8.7\times10^{-8}$ & $(1-2)\times10^{-8}$ & $\lesssim2\times10^{-8}$    \\
HVC & 90 & 3.0 & $30$   & $3$   & $\lesssim0.3$  &  & $2.5\times10^{-6}$  &
$2.5\times10^{-7}$ & $\lesssim2.5\times10^{-8}$ &  \\
LEC & 55 & 0.5 & $5$    & $\sim1$   & $\lesssim0.05$ &  & $2.5\times10^{-6}$  &
$\sim5\times10^{-7}$ & $\lesssim2.5\times10^{-8}$ & \\
            \noalign{\smallskip}
            \hline
         \end{tabular} 
\begin{list}{}{}
\item[$^{\mathrm{a}}$] Dust temperature associated with the continuum source
  against which the molecular absorption is produced. 
\item[$^{\mathrm{b}}$] Continuum optical depth of the continuum source against
  which the molecular absorption is produced. In the QC, it coincides with the
  region where the absorption is produced, but in the outflowing components
  (HVC and LEC) corresponds to the background continuum source (see Fig.~12 in
  GA17).
\item[$^{\mathrm{c}}$] From GA17.
\item[$^{\mathrm{d}}$] No significant upper limits in the outflowing components.
\item[$^{\mathrm{e}}$] The molecular abundances in the QC are inferred by assuming a
  gas-to-dust ratio of 100 and a mass absorption coefficient of
  $k_{\mathrm{abs}}=44.5$ cm$^2$/g of dust at 100 $\mu$m \citep{gon14b}. For
  the outflowing components, an OH abundance of $2.5\times10^{-6}$ relative to
  H nuclei is adopted (GA17), and all other abundances are derived from this
  normalization. 
\end{list}
\end{table*}

\subsubsection{Models for \hdop}

As noted in \S\ref{sec:h2opspec}, the \hdop\ lines in Mrk~231 are weak and
show absorption only at systemic velocities, with no evidence for blueshifted
wings. Figures~\ref{fith2o+}a-f show the predicted \hdop\ 
(and NH$_2$ in panel a) PACS lines of a 
model with \hdop\ column densities of $8.4\times10^{16}$ cm$^{-2}$ and
$3\times10^{15}$ cm$^{-2}$ in the QC and the HVC, respectively. These
\hdop\ columns are considered upper limits: while the model for the QC
approximately accounts for the $2_{21}-1_{10}$ $5/2-3/2$ 
and $3_{22}-2_{11}$ $5/2-3/2$ lines (Fig.~\ref{fith2o+}b,c),
it overpredicts the detected \hdop\ $3_{13}-2_{02}$ $7/2-5/2$ and
$5/2-3/2$ lines (Fig.~\ref{fith2o+}e) and the undetected
$3_{22}-2_{11}$ $7/2-5/2$ line (Fig.~\ref{fith2o+}d). Therefore, a
\hdop\ column of $(4.2-8.4)\times10^{16}$\,cm$^{-2}$ is favored for the QC, and
$N(\mathrm{H_2O^+})\lesssim3\times10^{15}$\,cm$^{-2}$ is obtained for the HVC
(Table~\ref{tab:col}). 

With these moderate columns, we emphasize the surprisingly strong emission
feature associated with the ground-state \hdop\ $1_{11}-0_{00}$ $1/2-1/2$ line 
in the SPIRE spectrum.  While the lack of apparent blueshifted absorption
in this line in consistent with the low \hdop\ column density found for the
HVC from the PACS lines, the emission feature would probably require either
very high volume densities in the region of the QC where the line is
generated, or formation pumping.

Unfortunately,
no collisional rates between \hdop\ and H atoms are available, so we have
attempted the following rough approach in order to check the consistency of
the \hdop\ identification and to establish a reference density based on an
(uncertain) assumption on the collisional excitation of \hdop. The \ohp--H
de-excitation collisional rates reported by \cite{liq16} between 
the excited \ohp\ levels and the ground-state $0_1$ state show little
dependence on gas temperature and follow the approximate
relationship $C_{u\rightarrow0_1}/g_{0_1}=8.5\times 10^{-10}\times \Delta E^{-0.6}$
  cm$^3$ s$^{-1}$, where $g_{0_1}=3$ is the degeneracy of the $0_1$ level
and $\Delta E$ is the energy of the upper level. This expression holds up to
$\Delta E>1000$ K, i.e. up to the highest energy levels considered by the
authors, $N=7$. We have simply assumed that the above expression, 
with a (rounded-off) multiplicative constant of $10^{-9}$ (rather than
$8.5\times 10^{-10}$) to include 
collisions with electrons, yields characteristic values for the de-excitation
rates between the O-bearing molecular ions in collisions with H atoms, and
applied it to the de-excitation rates of \hdop\ transitions to the ortho
($0_{00}$) and para ($1_{01}$ $3/2$ and $1/2$) ground levels. The excitation
rates are then 
calculated through detailed balance with an adopted gas temperature of 500 K.
For reference, the resulting collisional excitation rate from the
$0_{00}$ level to the $1_{11}$ $1/2$ level is $\approx8\times10^{-11}$ cm$^3$
s$^{-1}$. The collisional rates between excited levels are neglected. 

Our overall model for \hdop\ also includes in Fig.~\ref{fith2o+}g-h
calculations for H$_2$O and H$_2^{18}$O, the former based on the model
reported in \cite{gon10} by also including an outflowing component; results
for these species will be discussed in a future work. The main points relevant
for the present study are $(i)$ to approximately
reproduce the strength of the \hdop\ $1_{11}-0_{00}$ $1/2-1/2$ line with our
adopted collisional rates, a density
of $\mathrm{several}\times10^6$ cm$^{-3}$ in the QC is required, i.e. one
order of magnitude higher than that inferred from the \ohp\ ground-state
lines; $(ii)$ the model is roughly consistent with the lack of an emission
feature in the \hdop\ $1_{11}-0_{00}$ $3/2-1/2$ line (panel h), due to
cancellation by strong H$_2$O $1_{11}-0_{00}$ absorption in the outflow; 
$(iii)$ some contribution to the absorption wing observed in panel h) can be
attributed to \hdop, with the moderate column densities listed in
Table~\ref{tab:col}. 

We remark that the density inferred  for the QC from the ground-state
\hdop\ lines, $\mathrm{several}\times10^6$ cm$^{-3}$, should only be taken as a
qualitative confirmation of the results obtained for \ohp, 
but the absorption lines seen by PACS and the emission lines seen by SPIRE do
not sample the same regions (\S\ref{sec:chemqc}). Our subsequent analysis
(\S\ref{sec:chem}) is based on the column densities inferred from the
absorption lines, where the densities may be significantly lower. 


\subsubsection{Models for \htop}

As seen in \S\ref{sec:h3opspec}, little evidence is found in the PACS spectrum
of Mrk~231 for \htop, and an upper limit for the \htop\ column
density in the QC is estimated from the model shown in
Fig.~\ref{fith3o+}. With $N(\mathrm{H_3O^+})=8.4\times10^{16}$ cm$^{-2}$,
i.e. just the same column as used for \hdop\ in Fig.~\ref{fith2o+}, the
model accounts for some hints of absorption in Fig.~\ref{fith3o+}a-b, and is
also compatible with the undetected lines shown in Fig.~\ref{fith3o+}c--d. No
significant upper limits on the \htop\ column density in the outflowing
components could be estimated from the absence of line wings.

\subsection{Chemical models}
\label{sec:chem}

In the previous sections, radiative transfer models were applied to the
O-bearing molecular ions, enabling us to estimate the column densities and
abundances of each species in the three kinematic components that are needed
to account for the observed spectra (Table~\ref{tab:col}). Here we use the
results for the abundances, inferred from the strengths of the absorption
lines, as inputs to compare with chemical models, with 
the goal of estimating the ionization rate $\zeta$ of the molecular gas
components.

The chemical models presented here are the same as in GA13,
which were performed with the chemical code described in detail by 
\cite{bru09}, based on previous work by \cite{sta05} and \cite{dot02}. 
Briefly, the chemical models calculate the steady state abundances of 
all relevant species based on the UMIST06 reaction rates \citep{woo07}. The
gas with H nuclei density $n_{\mathrm{H}}$ and molecular 
fraction $f_{\mathrm{H_2}}$ is directly exposed to a cosmic ray and/or X-ray
flux that produces a total ionization rate per H nucleus $\zeta$ (including
secondary ionizations). We ignore any other external agent (e.g.\ dissociating
UV radiation, though internally generated UV radiation is included)  
and also the effect of negatively charged PAHs \citep{hol12}. 
Model results basically depend on
$\zeta/n_{\mathrm{H}}$, the molecular fraction $f_{\mathrm{H_2}}$, the gas
temperature ($T_{\mathrm{gas}}$), and the gas-phase oxygen abundance
($X(\mathrm{O})$).  We have generated a grid of models with representative
values of $T_{\mathrm{gas}}=150$, $550$, and 1000\,K, solar 
metallicities with $X(\mathrm{O})=3\times10^{-4}$, and $\zeta/n_{\mathrm{H}}$
in the range $5\times10^{-19}-5\times10^{-16}$ cm$^3$\,s$^{-1}$. Because of
uncertainties regarding H$_2$ formation on warm dust grains, and the
lack of equilibrium in the H$_2$ fraction that may affect mostly the outflow
components \citep{ric18} but also the QC, $f_{\mathrm{H_2}}$ is considered a
free parameter.

\subsubsection{The QC}
\label{sec:chemqc}

   \begin{figure*}
   \centering
   \includegraphics[width=18.0cm]{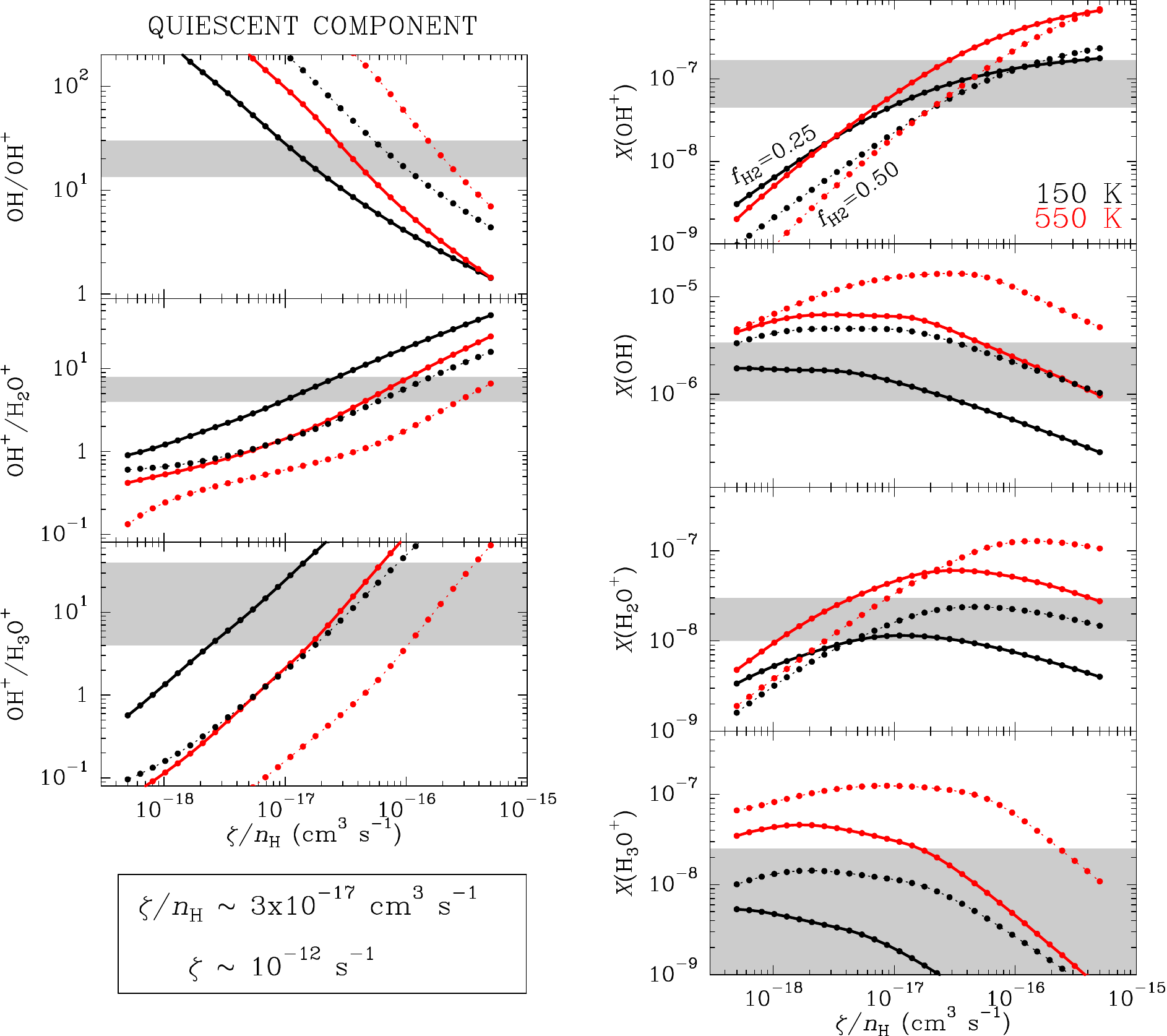}
   \caption{Chemical models (from GA13) for the QC. {\sl Left panels:} the
     predicted abundance ratios $\mathrm{OH/OH^+}$, $\mathrm{OH^+/H_2O^+}$,
     and $\mathrm{OH^+/H_3O^+}$, as a function of $\zeta/n_{\mathrm{H}}$. 
     {\sl Right panels:} the absolute abundances of the considered molecular
     species. Black and red lines indicate results for $T_{\mathrm{gas}}=150$
     and $550$\,K, respectively, while thick-solid and dotted lines use
     molecular fractions of $f_{\mathrm{H_2}}=0.25$ and $0.50$, respectively. The
     horizontal shaded regions indicate the range of values 
     favored by {\it Herschel} observations and our radiative transfer models
     for the QC. Our favored solution is in the range between the thick
     curves ($T_{\mathrm{gas}}=150-550$\,K, $f_{\mathrm{H_2}}=0.25$), for
     which $\zeta/n_{\mathrm{H}}\sim3\times10^{-17}$\,cm$^3$\,s$^{-1}$ (bottom
     left). 
   } 
    \label{chemqc}
    \end{figure*}

The predicted chemical model abundances of \ohp, OH, \hdop,
and \htop\ and their ratios are plotted in Fig.~\ref{chemqc} as a function of 
$\zeta/n_{\mathrm{H}}$ and compared with the
ranges allowed by the model fits to the observations of the QC.  
The most reliable observational values are the abundance
ratios shown in the left-hand panels, for which an uncertainty of a
factor $1.5$ is adopted.  Because \htop\ is not clearly detected in Mrk~231,
we adopt a large range of $4-40$ for the $\mathrm{OH^+/H_3O^+}$ ratio.
The absolute abundances are shown in
the right-hand panels, for which a higher uncertainty, of a factor of 2, is
adopted. Chemical models are shown for $T_{\mathrm{gas}}=150$ and 550\,K, and
for two fiducial values of the molecular fraction, $f_{\mathrm{H_2}}=0.25$ and
$0.50$.


Inspection of Fig.~\ref{chemqc} highlights the unique role of \ohp\ 
in probing the ionization rate of the molecular gas, as it is the only species
for which the abundance increases monotonically with $\zeta/n_{\mathrm{H}}$ up
to the most extreme values (for fixed $T_{\mathrm{gas}}$ and
$f_{\mathrm{H_2}}$). By contrast, the abundances of all other considered 
species decline with increasing $\zeta/n_{\mathrm{H}}$ above 
$\mathrm{several\,\times10^{-17}}$\,cm$^3$\,s$^{-1}$. 
In addition, Fig.~\ref{chemqc} indicates that results for both the absolute
abundances and abundance ratios are sensitive to both $f_{\mathrm{H_2}}$
and $T_{\mathrm{gas}}$. 
Once \ohp\ is formed, it can be destroyed either by generating \hdop, or
through the competing dissociative recombination that breaks the pathway to
generate \hdop\ and \htop. 
Increasing \tgas, the recombination rate decreases and the 
channel that produces \hdop\ and \htop\ 
(i.e. $\mathrm{OH^++H_2\rightarrow H_2O^++H}$ and  
$\mathrm{H_2O^++H_2\rightarrow H_3O^++H}$)
is reinforced, lowering the $\mathrm{OH^+/H_2O^+}$ and $\mathrm{OH^+/H_3O^+}$
ratios. In addition, the OH abundance is
boosted at high $T_{\mathrm{gas}}$ mainly because the activation barrier of
$\mathrm{O+H_2\rightarrow OH+H}$ is overcome \citep[e.g.][]{eli78}, thus
increasing the $\mathrm{OH/OH^+}$ ratio. On the other hand, the abundance of
\ohp\ relative to the other species also decreases with increasing 
$f_{\mathrm{H_2}}$, because the above reactions that generate 
\hdop\ and \htop\ proceed more efficiently.
Therefore, to explain the inferred abundance ratios (shaded regions in
Fig.~\ref{chemqc}), higher values of $\zeta/n_{\mathrm{H}}$ are required
if higher values for $f_{\mathrm{H_2}}$ and $T_{\mathrm{gas}}$ are adopted.

A range of physical conditions, and in particular in density,
$T_{\mathrm{gas}}$, $f_{\mathrm{H_2}}$, and $\zeta/n_{\mathrm{H}}$, will be
naturally present in the QC, though densities of $\mathrm{several}\times10^4$
cm$^{-3}$ are most likely prevalent \citep{mas15}.
At densities $\gtrsim5\times10^4$ cm$^{-3}$, the gas cools
efficiently and thermal balance (including only cosmic ray heating) indicates
that $T_{\mathrm{gas}}$ will be moderate even for high
$\zeta\sim5\times10^{-13}$\,s$^{-1}$ 
\citep[$<200$ K;][]{pap11,pap14}. Observationally, the CO lines from
$J_{\mathrm{up}}=5$ to 11 are fitted in Mrk~231 with $T_{\mathrm{gas}}\approx300$ K 
\citep{mas15}. We thus expect \tgas\ in the range between the 150 and 550 K
cases shown in Fig.~\ref{chemqc}, but probably closer to the
low-\tgas\ solution. 

To account for the inferred abundances and ratios, the required value of
$\zeta/n_{\mathrm{H}}$ increases by a factor of 
$\sim5$ if $f_{\mathrm{H_2}}$ is increased from $0.25$ to $0.50$
(Fig.~\ref{chemqc}).  
Nevertheless, 
the combination of very warm
$T_{\mathrm{gas}}=550$\,K and $f_{\mathrm{H_2}}\gtrsim0.50$ is ruled out, as
it grossly overpredicts the abundances of OH, \hdop, and \htop\ unless an
extremely high $\zeta/n_{\mathrm{H}}>10^{-16}$\,cm$^3$\,s$^{-1}$ is assumed,
but this would be hardly compatible with a high molecular fraction (GA13).
Indeed, the $f_{\mathrm{H_2}}=0.50$ case is only compatible with 
$T_{\mathrm{gas}}=150$ K, as it gives 
$\zeta/n_{\mathrm{H}}\gtrsim5\times10^{-17}$\,cm$^3$\,s$^{-1}$ that is still
consistent with a maximum $f_{\mathrm{H_2}}^{\mathrm max}\sim0.5$ (GA13).
On the other side, all molecular abundances start to decline
strongly at $f_{\mathrm{H_2}}<15$\% (not shown), and the molecular chemistry
is simply suppressed. Our favored solution is therefore 
$T_{\mathrm{gas}}=150-500$\,K and $f_{\mathrm{H_2}}=0.15-0.5$, with decreasing 
$T_{\mathrm{gas}}$ for increasing $f_{\mathrm{H_2}}$, for which we infer
$\zeta/n_{\mathrm{H}}\sim(3\pm2)\times10^{-17}$\,cm$^3$\,s$^{-1}$. 
The electron abundance is $\sim10^{-3}$. 

The quoted solution 
gives results compatible with the observational
constraints for the $\mathrm{OH/OH^+}$, $\mathrm{OH^+/H_2O^+}$, and
$\mathrm{OH^+/H_3O^+}$ ratios, and also nicely brackets the absolute
abundances estimated for \ohp, OH, \hdop, and \htop. In particular, the
predicted range for the abundance of OH ($(0.8-5)\times10^{-6}$)
is consistent with the value inferred in ULIRGs from the OH 35\,$\mu$m
transition \citep{sto17}. As will be shown in a future work,
however, these physical conditions underestimate the H$_2$O abundance. 
A range of physical values will be naturally present in the QC, and 
it is quite possible that H$_2$O preferentially forms and survives in warm,
high-density regions (i.e.\ $\zeta/n_{\mathrm{H}}$ lower than average). 
The strong submillimeter line of \hdop\ could also partially arise there.

We can now evaluate the role of formation pumping, relative to
collisional pumping, in the emission of strong ground-state lines
of \ohp. The number of photons generated per unit time and volume in the three
ground-state $1_J-0_1$ lines via formation pumping is
$\gamma^{1_J-0_1}_{\mathrm{pump}}\sim n_{\mathrm{H}}\, \zeta\, \epsilon_{\mathrm{OH^+}}$
(GA13), where $\epsilon_{\mathrm{OH^+}}$ is the efficiency with which
ionizations of hydrogen are transferred to \ohp\ production \citep{neu10}. 
We adopt a value $\epsilon_{\mathrm{OH^+}}=0.5$, which is an upper limit
because neutralization of H$^+$ on small grains or PAHs is expected to
decrease the efficiency of OH$^+$ formation \citep{hol12,ind12}.
Ignoring excitation by the radiation field, the corresponding
value generated through collisional excitation is 
$\gamma^{1_J-0_1}_{\mathrm{col}}\sim f_{0_1} \,n_{\mathrm{H}}^2 \,X(\mathrm{OH^+})\,
\sum_{u}C_{0_1\rightarrow u}$,
where $f_{0_1}$ is the fractional population in the ground $0_1$ level, and
the sum extends to all collisional rates from the $0_1$ level to
any other excited one \citep[yielding
  $1.55\times10^{-9}$\,cm$^3$\,s$^{-1}$ at 300\,K;][]{liq16}. The ratio of
these quantities is 
\begin{equation}
\frac{\gamma^{1_J-0_1}_{\mathrm{pump}}}{\gamma^{1_J-0_1}_{\mathrm{col}}} \sim
\frac{(\zeta/n_{\mathrm{H}}) \,\epsilon_{\mathrm{OH^+}}}{
f_{0_1}\,X(\mathrm{OH^+})\,\sum_{u}C_{0_1\rightarrow u}}
\sim \frac{0.1}{f_{0_1}}\times 
\frac{\zeta/n_{\mathrm{H}}}{3\times10^{-17}\,\mathrm{cm^3\,s^{-1}}}
\end{equation}
where $X(\mathrm{OH^+})=10^{-7}$ is adopted (Fig.~\ref{chemqc}).
In our best-fit models, $f_{0_1}$ lies in the range $\sim0.2-0.6$, and thus 
collisional excitation appears to dominate\footnote{We have also
    evaluated the expected flux arising from formation pumping in the three 
ground-state \ohp\ lines using Appendix~A in GA13 and including also the
outflowing components, with the prediction that $\lesssim30$\,Jy\,km\,s$^{-1}$
are expected, much lower than the observed fluxes
(Table~\ref{tab:spirefluxes}).}. 


In summary, the combination of {\it Herschel} observations, radiative transfer
models, and chemical models, favor
$\zeta/n_{\mathrm{H}}\sim(3\pm2)\times10^{-17}$\,cm$^3$\,s$^{-1}$ in the
Kl\"ockner et al. torus. This is similar to, though likely somewhat higher
than, the value inferred in the nuclear 
region of Arp~220 (GA13), and indeed the continuum-normalized absorption
at systemic velocities in the excited \ohp\ and \hdop\ lines are similar in
both sources. One obvious difference, however, is the high densities inferred
in Mrk~231 from the ground-state lines of \ohp, 
$n_{\mathrm{H}}\sim2\times10^5$\,cm$^{-3}$ (\S\ref{oh+models}), pointing to a
very high value of $\zeta$ in the nuclear region. The caveat, however, is that
the column densities and abundances are primarily derived from the (excited)
absorption lines that are formed in front of the continuum source, while the
above density is inferred from the (ground-state) emission lines, which do not
have the above requirement and are primarily formed in a different region
of the torus. In fact, while the peak absorption of the excited
\ohp\ $2_J-1_{J'}$ lines is blueshifted, the peak emission of the ground-state
\ohp\ $1_J-0_{J'}$ lines is redshifted (Fig.~\ref{fitoh+}), unambiguosly
indicating distinct formation regions for the absorption and emission
lines. Observations of CO $2-1$ and $3-2$ show that the redshifted line wing
emission is stronger than that of the blueshifted wing \citep{fer15}.
Therefore, the density in the environments where the absorption lines 
are generated may be different, and probably lower than estimated from the
emission lines. Nevertheless, the warm CO SLED observed in the source
indicates that the overall density in the nuclear region of Mrk~231 is at
least $\mathrm{several}\,\times10^4$\,cm$^{-3}$ \citep{wer10,mas15}, from
which we conservatively estimate $\zeta\sim10^{-12}$\,s$^{-1}$ with an
uncertainty of up to a factor $\sim3$. This value is still $>10^3$ times
the mean value found in the Milky Way \citep{ind15},
$>10$ times the value estimated in a spiral galaxy at $z=0.89$
  \citep{mul16}, and $\gtrsim10$ times the maximum values found in the
  Galactic center by \cite{ind15}.

\subsubsection{The outflowing components}

   \begin{figure*}
   \centering
   \includegraphics[width=18.0cm]{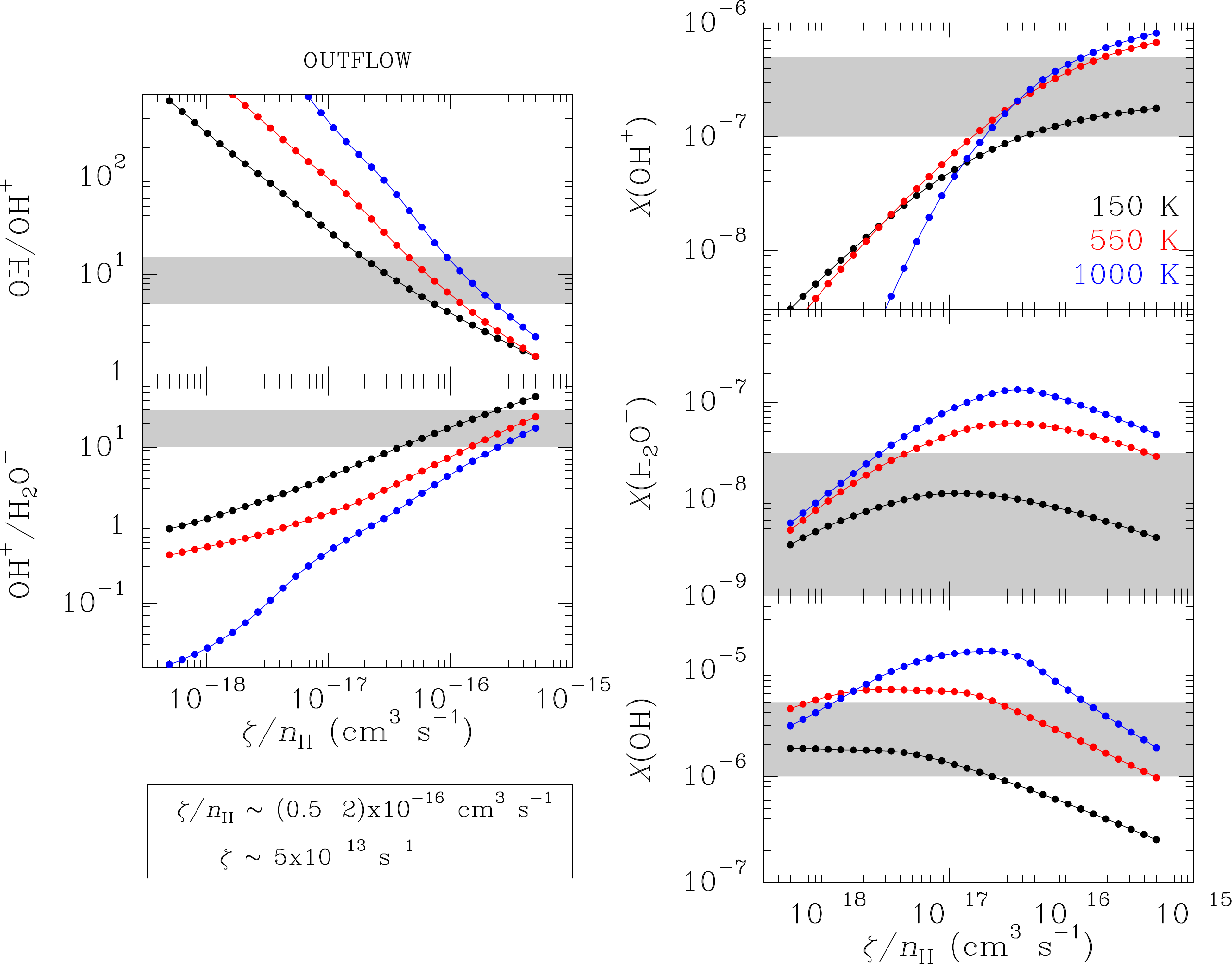}
   \caption{As Fig.~\ref{chemqc}, but for the outflowing components. 
     {\sl Left panels:} 
     the predicted abundance ratios $\mathrm{OH/OH^+}$ and
     $\mathrm{OH^+/H_2O^+}$ as a function of $\zeta/n_{\mathrm{H}}$. 
     {\sl Right panels:} the absolute
     abundances OH, \ohp, and \hdop. Model predictions are shown for 
     $T_{\mathrm{gas}}=150$ (in black), 550 (red), and 1000\,K (blue), and
     the molecular fraction is $f_{\mathrm{H_2}}=0.25$ in all models.
     At the bottom left corner, the favored values for both  
     $\zeta/n_{\mathrm{H}}$ and $\zeta$ for the outflow are indicated.
   } 
    \label{chemof}
    \end{figure*}

Because of the limited observational constraints for the outflowing gas, 
we model the HVC and the LEC with a single chemical model and
consider only the abundances of OH, \ohp, and \hdop\  
(Fig.~\ref{chemof}). In the outflow, a very low $\mathrm{OH/OH^+}\sim10$
abundance ratio is inferred (Table~\ref{tab:col}), which points to values 
of $\zeta/n_{\mathrm{H}}$ even higher than in the QC. 

To constrain $f_{\mathrm{H_2}}$ and the density $n_{\mathrm{H}}$, we follow
\cite{ric18,ric17} 
to characterize the physical conditions in the high-velocity (partially)
molecular outflows driven by luminous AGN. In their thermo-chemical 3D models,
the ISM gas is swept out in a forward shock driven by a hot shocked bubble,
attaining high $T_{\mathrm{gas}}\sim10^7$ K where all molecules are
destroyed. The molecular gas reforms downstream, and one key
result of these models is that there is a range in the post-shock density
($n_{\mathrm{H}}$, hence also in the pre-shock density $n_{\mathrm{H0}}$)
that allows for high-velocities of the outflowing molecular gas: if 
$n_{\mathrm{H}}$ is too low, the post-shock gas cannot cool in time to
form molecules, while if $n_{\mathrm{H0}}$ is too high, the post-shock
gas does not approach the observed high velocities.  
Values of $n_{\mathrm{H0}}$ around $10$ cm$^{-3}$ and post-shock gas
densities around $\mathrm{a\,few}\times10^3-10^4$ cm$^{-3}$, appear to 
be an appropriate compromise between quick enough H$_2$ formation
and high outflowing velocities \citep{ric17}, and these are indeed the
typical densities we found in our best fit model for the HVC  (GA17). In
addition, the conversion to molecular gas is far from complete, with
$f_{\mathrm{H_2}}\sim0.25$ being favored for high velocity outflows 
\citep{ric18}. 

Chemical models for the abundances and ratios of OH, \ohp, and \hdop\ 
in Fig.~\ref{chemof} therefore use $f_{\mathrm{H_2}}\sim0.25$ and, because the
gas is very warm at the ``knee'' of the H$_2$ formation, models for 
$T_{\mathrm{gas}}=1000$\,K are also considered. Results for the abundance
ratios are then consistent with
$\zeta/n_{\mathrm{H}}\sim(0.5-2)\times10^{-16}$\,cm$^3$\,s$^{-1}$. 
\cite{pap11} show that for an ionization
rate of $5\times10^{-13}$\,s$^{-1}$ (due to cosmic rays, see \S\ref{crdr})
and gas densities of $5\times10^{3}$\,cm$^{-3}$, \tgas\ is well above 200 K
even at high visual extinctions, and this is a lower limit because of
additional heating sources in the post-shock gas. Therefore, the high 
$T_{\mathrm{gas}}=550-1000$\,K solutions are favored for the outflowing gas,
which yield $\zeta/n_{\mathrm{H}}\sim(1-2)\times10^{-16}$\,cm$^3$\,s$^{-1}$ and
$\zeta\sim5\times10^{-13}$\,s$^{-1}$. This solution yields absolute
abundances of \ohp\ and \hdop\ above the observationally determined ranges,
probably meaning that only a fraction of the outflowing column of gas is 
subject to the above extreme conditions. For the HVC, where 
$N(\mathrm{OH^+})\sim3\times10^{16}$ \cmd\ (Table~\ref{tab:col}), the
corresponding hydrogen column is $4.3\times10^{22}$ \cmd\ for 
$X(\mathrm{OH^+})\sim7\times10^{-7}$, i.e. less than half the total
column of the HVC component. 
As the molecular gas cools and its density increases (with the
  concomitant decrease of $\zeta/n_{\mathrm{H}}$), the \ohp\ abundance is
  expected to decrease (Fig.~\ref{chemof}) and thus we expect that
  \ohp\ primarily probes the most extreme, partially molecular environments
  associated with high $T_{\mathrm{gas}}$ and $\zeta/n_{\mathrm{H}}$
  --including the molecular formation region of the post-shock gas. 


One caveat to the favored solution concerns the shallow H {\sc i} 21cm
absorption found by \cite{mor16} in the outflow of Mrk~231. For
$f_{\mathrm{H_2}}=0.25$, we expect a column 
$N(\textrm{H {\sc i}})\sim3.2\times10^{22}$ \cmd\ in the HVC, and about half
of this column is at blueshifted velocities $|v|>500$ \kms. While this is
still consistent with the observed H {\sc i} 21cm absorption at blueshifted
velocities if $T_{\mathrm{spin}}\gtrsim1000$ K, the model would overestimate
the observed absorption if $T_{\mathrm{spin}}$ were significantly lower. On
the other hand, at the expected densities $>10^3$\,cm$^{-3}$ the H {\sc i}
21cm line is thermalized and $T_{\mathrm{spin}}=T_{\mathrm{gas}}$
\citep{mor16}. Therefore, 
the \ohp\ and H {\sc i} 21cm absorption at high blueshifted velocities are
expected to arise in the same outflowing component as long as \tgas\ is high. 
Nevertheless, the shallow H {\sc i} absorption may be better explained if
a significant fraction of the synchrotron 21cm continuum is generated in
the forward shock \citep[][see also \S\ref{crgen}]{fau12,nim15,liu17},
i.e. ahead of the neutral and molecular outflowing gas, in such a way that
only a fraction of the radio continuum is background relative to H {\sc i}
\citep[see Figs.~1 in][]{fau12,zub14,ric17}.

\section{Discussion}
\label{sec:discussion}

\subsection{Summary of radiative transfer and chemical models}

A picture of both the torus around the AGN of Mrk~231 and the
outflowing molecular gas emerges from the combination of {\it Herschel}
spectroscopic observations in the far-IR, radiative transfer models to infer
the molecular column densities and abundances, and chemical
models that enable an estimation of the physical parameters responsible for
the observed chemistry.  

For the torus (denoted as QC above because it is responsible for the
relatively narrow spectral features observed near systemic velocities), we
find high column densities of dust, OH and \ohp, and relatively high
$\mathrm{OH^+/H_2O^+}\sim4-8$ and $\mathrm{OH^+/H_3O^+}>4$ ratios. 
The densities are high, but molecular
fractions $f_{\mathrm{H_2}}\lesssim50$\% are inferred together with high
ionization rates ($\zeta\sim10^{-12}$\,s$^{-1}$). 
Due to the extreme conditions around the AGN, which is expected to produce
high quantities of cosmic rays, only molecular gas at high densities, lowering
the $\zeta/n_{\mathrm{H}}$ ratio, can survive.  All gas
with densities below $\textrm{a few}\times10^4$ cm$^{-3}$ will be ionized and
ultimately dispersed.  

The molecular outflow observed in OH shows, on the other hand, even higher
\ohp\ columns relative to both OH and \hdop. This appears to point to even
higher values of $\zeta/n_{\mathrm{H}}\sim10^{-16}$ cm$^3$ s$^{-1}$, 
most probably due to densities lower than in the QC, and low molecular
fractions. Under these conditions the C$^+$ abundance is high
($\gtrsim10^{-4}$), which may explain the approximately 1:1 relationship in
the gas masses derived from OH and [C {\sc ii}] \citep{jan16}.
Nevertheless, a range of conditions is expected because efficient formation
of CO and HCN requires lower $\zeta/n_{\mathrm{H}}$ and higher
$f_{\mathrm{H_2}}$. Indeed, evidence for chemical differentiation has been 
found in the molecular outflow of Mrk~231 \citep{lin16}.

Quantitatively, the main caveat to our analysis concerns the absolute
molecular abundances in the outflow, for which we have adopted a reference,
fiducial value for OH of $2.5\times10^{-6}$ (GA17) -similar to that inferred
at systemic velocities from several transitions by assuming a gas-to-dust
ratio of $\sim100$ by mass \citep{gon12,sto17}. We emphasize that
the adopted OH abundance is relative to H nuclei in the
outflow, regardless of whether it is in atomic or molecular form
\citep{ric18}. In this way, we 
argue from the present study that the abundances of OH and \ohp\ adopted for
the outflow are most probably accurate to within a factor $\sim2.5$. If
$X(\mathrm{OH})$ were much higher than $2.5\times10^{-6}$, $X(\mathrm{OH^+})$
would also be much higher than $2.5\times10^{-7}$, which would be hard to
account for with current chemical models. On the other side, if
$X(\mathrm{OH})$ were much lower than $2.5\times10^{-6}$, the energetics
associated with the outflow (which scale as $X(\mathrm{OH})^{-1}$) would be
very hard to reconcile with detailed feedback models \citep{ric18,ric17}.

   \begin{figure}
   \centering
   \includegraphics[width=8.2cm]{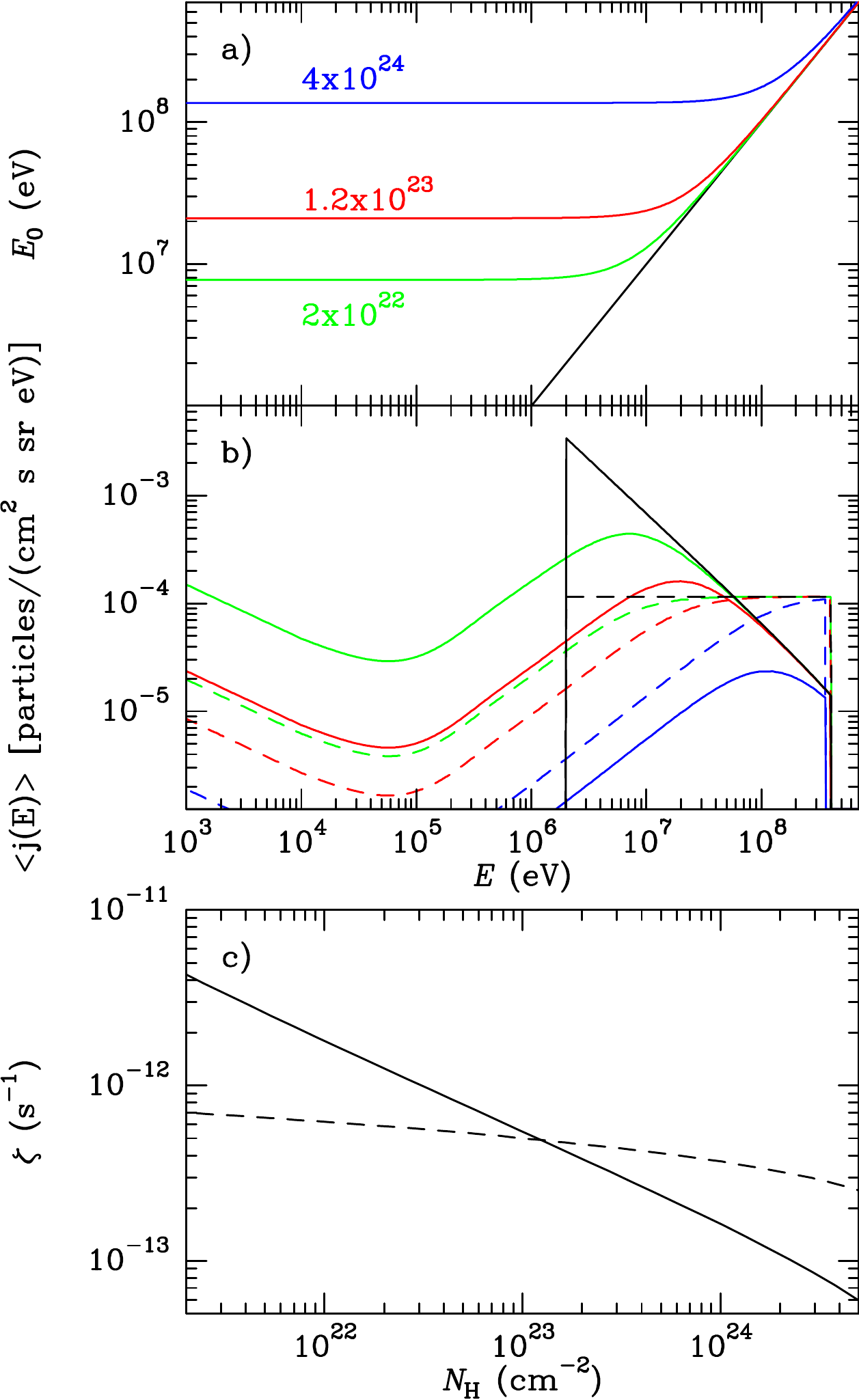}
   \caption{a) The CR proton energy $E$ after crossing a column
     $N_{\mathrm{H}}$ in terms of the incident CR energy at the cloud surface
     $E_0$ (eq.~\ref{eq:e0e}). Columns of $N_{\mathrm{H}}=2\times10^{22}$,
     $1.2\times10^{23}$, 
     and $4\times10^{24}$ cm$^{-2}$, corresponding to those inferred for the
     LEC, HVC, and QC components, are considered. There are
     thresholds in $E_0$ of 8, 21, and 140 MeV, respectively, which are the
     minimum incident CR energies that can penetrate 
     the corresponding columns. The black line indicates $E_0=E$. 
     b) The black lines show our adopted angle-averaged incident
     CR proton spectra (i.e. $<j_0(E_0)>$). We have used a power-law spectrum
     (solid line), with $<j_0>\propto p^{-2}$, and a flat distribution
     (dashed line). The colored curves show the propagated 
     spectra ($<j(E,N_{\mathrm{H}})>$) after crossing the three columns of gas
     considered in a) (solid and dashed lines for the power-law and flat
     distributions, respectively). c) The inferred ionization  
     rate at the center of a spherical cloud as a function of its column
     density, for both the power-law (solid) and flat (dashed) incident
     spectra. 
   } 
    \label{crs}
    \end{figure}

\subsection{A giant Cosmic Ray Dominated Region in the nucleus of Mrk~231}
\label{crdr}

What mechanism is responsible for the high ionization rates that we have
derived? 
{\it NuSTAR} observations of Mrk~231 have revealed an intrinsically weak 
source of X-rays, with a luminosity $L_X(2-10\,\mathrm{keV})=4\times10^{42}$
erg\ s$^{-1}$ \citep{ten14}. Assuming that the X-rays originate in the
AGN, the unattenuated X-ray flux at a distance of $r_{\mathrm{pc}}$ pc is
$F_X(2-10\,\mathrm{keV})=3.3\times(r_{\mathrm{pc}}/100)^{-2}$ 
erg\ s$^{-1}$ cm$^{-2}$. Following \cite{mal96}, the ionization rate produced
by the X-rays is given by
$\zeta(\mathrm{s^{-1}})\sim10^{-13}\,F_X/N_{22}^{0.9}$, where
$N_{22}=N/10^{22}\,\mathrm{cm^{-2}}$ and $N$ is the hydrogen column density
attenuating the X-ray flux. From these estimates, and given the extremely high
values of $\zeta\sim 5\times10^{-13}$ s$^{-1}$ we infer at $\gtrsim100$ pc and
affecting columns of gas $\gtrsim2\times10^{22}$ cm$^{-2}$ (Table~2 in GA17),
we find that the X-rays from the AGN are probably unable to account for the
{\it Herschel}/PACS observations 
of molecular ions in Mrk~231. Even if a putative luminous X-ray source were 
located at or close to the location of the AGN and strongly attenuated by
foreground gas, the 
excited lines of \ohp\ are observed in absorption against a continuum that is
optically thick even in the far-IR, which would protect the outflowing gas
from strong X-ray ionization. An alternative is that the observed X-ray
emission originates in the forward shock \citep{nim15}, in which case the
source of X-rays would be more spatially linked to the molecular outflow, 
although the mentioned X-ray attenuation would still apply.
We thus resort in the following to the potential
role played by CRs.

An ultra-fast outflow (UFO) with a velocity of $\sim2\times10^4$ km s$^{-1}$
has been recently identified in the {\it Chandra} and {\it NuSTAR} X-ray data
of Mrk~231 \citep{fer15}. The authors estimate that the UFO covering
fraction is consistent with unity, with mass outflow and energy rates in the
range $\dot{M}_{\mathrm{UFO}}=0.3-2.1$ M$_{\odot}$ yr$^{-1}$ and 
$\dot{E}_{\mathrm{UFO}}=(0.38-2.7)\times10^{44}$ erg s$^{-1}$, respectively. 
The wind mechanical power is significantly lower than that inferred in
IRAS~F11119$+$3257, where a molecular outflow is also detected
\citep{tom15,vei17}. 
The UFO in Mrk~231 is detected during an epoch of low radio emission
(i.e. suppressed jet activity) but undetected in epochs of high radio
emission, suggesting high variablity and possible jet-wind mutual exclusion
\citep{rey17}. The UFO, jet activity \citep[e.g.][]{ulv99}, BAL wind
\citep[e.g.][]{vei13b,vei16}, and molecular outflow suggest the accompanying
generation of CRs \citep{fau12,nim15,liu17} that can permeate and ionize the
molecular gas.

We attempt here to evaluate the energetic requirements of the CR field in
Mrk~231 that is responsible for the inferred ionization rates. The energy loss
of a CR proton that has propagated through an atomic cloud with column
$N_{\mathrm{H}}=2\times10^{22}$, $1.2\times10^{23}$, and $4\times10^{24}$
cm$^{-2}$ is shown in Fig.~\ref{crs}a. We have used the
{\it continuous-slowing-down-approximation} \citep{pad09} and have calculated
$E_0$ numerically for a given final energy $E$ according to
\begin{equation}
\int_{E}^{E_0} \frac{dE}{L_k(E)}=N_{\mathrm{H}},
\label{eq:e0e}
\end{equation}
where $L_k(E)=-dE/dN_{\mathrm{H}}$ is the energy loss function taken from the
NIST
database\footnote{https://www.nist.gov/pml/stopping-power-range-tables-electrons-protons-and-helium-ions. We
ignore the energy loss by pion production, relevant at proton energies above 1
GeV.}. Figure~\ref{crs}a
indicates that incident proton energies $E_0>8$, 21, and 140 MeV are required
to penetrate the quoted column densities. These energies are significantly
higher than the 2 MeV protons associated with the UFO, 
and we thus require a different particle (CR) field that can
penetrate the high columns of gas associated with the kinematic components
in Mrk~231 (QC, HVC, LEC).

Given the threshold energies of Fig.~\ref{crs}a, and the decreasing Bethe
cross section for primary ionization of atomic hydrogen 
($\sigma_{\mathrm{ion}}\sim E^{-0.88}$ between 2 and 100 MeV), the ionization
is expected to be mainly caused by CRs with incident energies between 
$E_0\sim8$ and $\mathrm{several}\,\times100$ MeV. 
Since we aim to check the ionization 
rate as a function of column density, we have considered CR protons with
energies between 2 and 400 MeV, and adopted two incident spectra: a power-law
distribution with $\sim p^{-2}$ (solid black line in Fig.~\ref{crs}b), and a
flat distribution (dashed black line).
The normalization of each distribution is a free parameter that we vary in
order to obtain appropriate values of $\zeta$. The propagated spectra after
crossing the characteristic column densities quoted above are also shown in
Fig.~\ref{crs}b with colored lines. We have again used the method outlined by
\cite{pad09}, i.e.
\begin{equation}
j(E,N_{\mathrm{H}})= j_0(E_0)\times\frac{L_k(E_0)}{L_k(E)},
\end{equation}
where $j_0(E_0)$ is the incident spectrum and $j(E,N_{\mathrm{H}})$ is the
propagated spectrum after crossing a column density of $N_{\mathrm{H}}$. 
The propagated spectra have a minimum at $\sim70$ keV because there is a
maximum in $L_k(E)$ at this energy \citep{pad09}.

The ionization rates at the cloud center, shown in Fig.~\ref{crs}c as a
function of $N_{\mathrm{H}}$, are calculated as \citep[e.g.][]{spi68} 
\begin{equation}
\zeta = \frac{5}{3} (1+G_{10})\, 4\pi \int dE <j(E,N_{\mathrm{H}})>\, \sigma(E)  
\end{equation}
where $\sigma$ is the Bethe ionization cross section, the factor $5/3$
accounts for secondary ionizations, and $G_{10}=0.5$ simulates the
contribution to the ionization by nuclei heavier than protons, assuming solar
metallicities \citep[see Appendix in][]{ind09}. With the normalizations
showed in Fig.~\ref{crs}b, the values of $\zeta$ are 
$\sim5\times10^{-13}$ s$^{-1}$ at 
$N_{\mathrm{H}}=1.2\times10^{23}$ cm$^{-2}$ (Fig.~\ref{crs}c), thus
accounting for the ionization rate inferred in the outflowing HVC component. 
CRs with energies above 400 MeV would be required to reproduce
$\zeta\sim10^{-12}$ s$^{-1}$ in the QC, 
although CRs originating from supernovae within the QC could also provide an
important contribution.

The incident angle-averaged CR spectrum shown in Fig.~\ref{crs}b is related to
the mass outflow rate, momentum flux, and energy flux in CRs, as
\begin{eqnarray}
\dot{M}_{\mathrm{CR}} & = & (4\pi)^2 R^2 f_c \, \mu \, m_p \int dE <j_0(E)> \\
\dot{p}_{\mathrm{CR}} & = & (4\pi)^2 R^2 f_c \, \mu \int dE <j_0(E)> p(E) \\
\dot{E}_{\mathrm{CR}} & = & (4\pi)^2 R^2 f_c \, \mu \int dE <j_0(E)> E,
\label{dotecr}
\end{eqnarray}
where $m_p$ is the proton mass, $\mu=1.4$ corrects for the mass of heavier
nuclei, $p(E)$ is the proton momentum correponding to
kinetic energy $E$, and $R$ is the distance to the AGN. The inclusion of the
factor $f_c$, the covering factor of the outflow (GA17), makes our estimate
conservative, as we only count the CRs incident on the molecular region traced
by \ohp\ (the aligned-momenta approach, GA17). 
Using $R=170$ pc and $f_c=0.18$, appropriate for the HVC (GA17), we obtain 
$\dot{M}_{\mathrm{CR}}=0.010-0.013$\,\Msun\,yr$^{-1}$, 
$\dot{p}_{\mathrm{CR}}=(0.6-1.6)\times10^{34}$\,dyn, 
and $\dot{E}_{\mathrm{CR}}=(0.44-1.7)\times10^{44}$\,erg\,s$^{-1}$ for the
power-law and flat distributions, respectively. 
Even higher values are obtained for the LEC with $R=600$ pc and $f_c=0.26$,
but the more moderate column in this component means that protons
in the high-velocity wing of the thermal distribution of the hot bubble at
$\sim10^{10}$ K, as predicted by models of energy-conserving outflows 
\citep[e.g.][]{kin11,zub12,fau12,ric18}, could mix with the outflowing ISM
layer and contribute significantly to increase $\zeta$. The quoted values are
{\it instantaneous} \citep[GA17,][]{vei17}, and the time-averaged values are a
factor $\sim4$ smaller.

\subsection{The generation of CRs and the forward shock in Mrk~231}
\label{crgen}

While the inferred $\dot{M}_{\mathrm{CR}}\sim0.01$ \Msun yr$^{-1}$ is
significantly lower than $\dot{M}_{\mathrm{UFO}}$, it is worth noting that
$\dot{E}_{\mathrm{CR}}\sim10^{44}$ erg s$^{-1}$ is similar to 
both $\dot{E}_{\mathrm{UFO}}$ and the time-averaged energy flux associated
with the molecular outflow \citep[$\sim10^{44}$ erg s$^{-1}$;][GA17]{fer15},
and equivalent to $\sim1$\% of the AGN luminosity ($\sim10^{46}$ erg
s$^{-1}$). CRs are not thermalized into a hot bubble, but penetrate into the
gas producing ionizations and depositing momentum as well. 
Although $\dot{p}_{\mathrm{CR}}$ is a lower limit as it neglects CRs magnetic
mirroring \citep{pad11}, the estimated momentum rate of the CR field still
falls short of accounting for the inferred momentum rate of the molecular
outflow ($\sim2.5\times10^{36}$ dyn for the HVC and LEC together, GA17). 

Multitransition analysis of the OH doublets yields estimated sizes (radii)
for the molecular outflow components in Mrk~231 of 170\,pc (HVC) and 600\,pc
(LEC, GA17). These sizes are similar to those measured in the radio 
continuum for the inner and outer disk-like components of the source,
respectively \citep{car98,tay99}. 
It is thus tempting to relate the CR proton field
responsible for the molecular ionization of the outflowing gas to the
relativistic electrons that generate the disk-like synchrotron radio 
emission. Specifically, we evaluate whether both the CR protons and electrons
can be accelerated in the forward shock, presumably driven by the AGN 
though with possible contribution by the starburst, that
sweeps out the ISM generating the 
observed molecular outflow. Following \cite{fau12} and \cite{nim15}, the
predicted synchrotron emission associated with the HVC is given by 
$\nu L_{\nu} \sim 5.4\times10^{-6} \epsilon_{-2} L_{\mathrm{AGN}} 
(L_{\mathrm{kin}}/0.027L_{\mathrm{AGN}})$, where
$\epsilon=10^{-2}\epsilon_{-2}$ is the fraction of the forward shock energy
flux that goes into relativistic electrons, and we have normalized the
instantaneous mechanical power of the HVC in Mrk~231 to $2.7$\% of
$L_{\mathrm{AGN}}$  (GA17, assuming that it is a factor of $\sim4$ higher than
the time-averaged value). Using $\epsilon_{-2}=1$ and
$L_{\mathrm{AGN}}=8.6\times10^{45}$\,erg\,s$^{-1}$, the 
predicted monochromatic luminosity at 1.4 GHz is
$L_{\mathrm{1.4\,GHz}}\sim3\times10^{31}$\,erg\,s$^{-1}$\,Hz$^{-1}$,  
while the observed 1.4 GHz luminosity from the inner disk of Mrk~231 is 
$5.3\times10^{30}$\,erg\,s$^{-1}$\,Hz$^{-1}$ \citep{car98}. 
Despite inverse-Compton losses may significantly decrease the predicted 1.4
GHz luminosity, the excess of a factor $\sim5$ relative to observations
appears to indicate that there is enough
energy flux in the forward shock associated with the molecular outflow to
account for the observed disk-like radio emission \citep[for the ionized
  phase, see][]{zak14}. 

On the other hand, the CR
proton luminosity (i.e. the energy flux associated with the CRs) is estimated
as $\dot{E}_{\mathrm{CR}}\sim 2.7\times10^{-3} (\epsilon_{\mathrm{nt}}/0.1) 
L_{\mathrm{AGN}} (L_{\mathrm{kin}}/0.027L_{\mathrm{AGN}})$, where 
$\epsilon_{\mathrm{nt}}\sim0.1$ is the assumed fraction of the forward energy
flux that is converted into non-thermal energy flux of CR protons
\citep{liu17}. The above expression yields 
$\dot{E}_{\mathrm{CR}}\sim 2\times10^{43}$\,erg\,s$^{-1}$, at least a factor of
$\sim2$ below our estimate from eq.~(\ref{dotecr}). The discrepancy may be due
to a higher value of $\epsilon_{\mathrm{nt}}$ for the low-energy CRs causing
the molecular ionization. In addition, and according to first-order Fermi
acceleration, CRs gain energy by crossing and recrossing the shock front back
and forth \citep[e.g.][]{bel13}, and the observed ionization rate 
may be caused by diffusion of these accelerating CRs into the outflowing
molecular gas downstream. The process may be very efficient, as a given CR 
crosses the shock many times.


\section{Conclusions}
\label{sec:conclusions}

The main findings and conclusions of this study are: \\
1- {\it Herschel}/PACS spectroscopic observations of Mrk~231 show absorption
in the OH$^+$ excited rotational lines at both systemic and
blueshifted velocities up to $\sim-1000$ \kms. OH$^+$ is thereby an excellent
tracer of the molecular outflow. \\
2- The OH$^+$ $2_2-1_1$ and $2_1-1_0$ fine-structure lines at $\sim150$ $\mu$m
show blueshifted absorption wings very similar in shape to those observed in
the OH doublets at 84 and 65 $\mu$m, indicating that both species share
similar outflow regions. \\ 
3- {\it Herschel}/PACS observations of the excited \hdop\ and
\htop\ rotational lines only show absorption, if any, at systemic velocities,
with no evidence for blueshifted wings. Most \htop\ rotational lines are
undetected. \\
4- {\it Herschel}/SPIRE observations show P-Cygni profiles in the ground-state
\ohp\ lines, with strong redshifted emission features. The ground-state 
\hdop\ $1_{11}-0_{00}\,1/2-1/2$ line shows strong emission above the
continuum, but no hints of blueshifted absorption. \\
5- Radiative transfer models, similar to those previously reported for OH
(GA17), have been applied to \ohp, \hdop\ and \htop. At systemic velocities,
probing the nuclear torus of $\sim100$ pc scale around the AGN, we find column
density ratios of $\mathrm{OH/\ohp}\sim20$, $\mathrm{\ohp/\hdop}\sim4-8$, and
$\mathrm{\ohp/\htop}\gtrsim4$. For the outflowing gas, 
$\mathrm{OH/\ohp}\sim10$ and $\mathrm{\ohp/\hdop}\gtrsim10$. The abundance of
\ohp\ relative to H nuclei is estimated to be high, $>10^{-7}$, in both
components. \\ 
6- Chemical models are used to predict the ratios and absolute values of the
abundances of OH, \ohp, \hdop, and \htop, which are compared with the inferred
values to estimate the ionization rate $\zeta$ of the molecular gas. We
estimate $\zeta\sim10^{-12}$ and $\sim5\times10^{-13}$\,s$^{-1}$ in the
nuclear torus and outflowing gas, respectively. The molecular fraction is
expected to be low, $f_{\mathrm{H_2}}<0.5$. \\
7- The inferred high ionization rates, of order $10^4$ times those inferred in
the Milky Way, are hard to explain by the relatively weak emission of Mrk~231
in X-rays, and therefore low-energy ($10-400$ MeV) cosmic rays are proposed to
play a primary role on the ionization of the molecular gas. Accounting for the
energy loss as CRs travel through the (partially) molecular gas, we
estimate a mass outflow rate and energy flux in low-energy CRs of
$\dot{M}_{\mathrm{CR}}\sim0.01$\,\Msun\,yr$^{-1}$ and 
$\dot{E}_{\mathrm{CR}}\sim10^{44}$\,erg\,s$^{-1}$. \\
8- Diffusion of CRs downstream into the outflowing molecular gas, as they are
accelerated through repeated crossings of the forward shock, may be an
efficient way of accounting for the ionization rate of the outflowing gas. The
forward shock associated with the molecular outflow in Mrk~231 also has enough
energy to account for the disk-like synchrotron emission generated by
relativistic electrons.

\acknowledgments
We thank Alexander Richings and Claude-Andr\'e Faucher-Gigu\`ere
for useful discussions on models of
energy-conserving outflows, and Karen Yang and Ke Fang for commenting on the
problem of CR propagation. 
E.GA is grateful for the warm hospitality of the Harvard-Smithsonian
Center for Astrophysics, where most of the present study was carried out.
PACS was developed by a consortium of institutes
led by MPE (Germany) and including UVIE (Austria); KU Leuven, CSL, IMEC
(Belgium); CEA, LAM (France); MPIA (Germany); 
INAFIFSI/OAA/OAP/OAT, LENS, SISSA (Italy); IAC (Spain). This development
has been supported by the funding agencies BMVIT (Austria), ESA-PRODEX
(Belgium), CEA/CNES (France), DLR (Germany), ASI/INAF (Italy), and
CICYT/MCYT (Spain). E.GA is a Research Associate at the Harvard-Smithsonian
Center for Astrophysics, and thanks the Spanish 
Ministerio de Econom\'{\i}a y Competitividad for support under projects
FIS2012-39162-C06-01 and  ESP2015-65597-C4-1-R. E.GA and H.A.S. 
thank NASA grant ADAP NNX15AE56G. Basic
research in IR astronomy at NRL is funded by 
the US ONR; J.F. also acknowledges support from the NHSC. 
S.V. thanks NASA for partial support of this research via Research Support
Agreement RSA 1427277, support from a Senior NPP Award from NASA, and
his host institution, the Goddard Space Flight Center.
This research has made use of NASA's Astrophysics Data System (ADS)
and of GILDAS software (http://www.iram.fr/IRAMFR/GILDAS).

\facilities{{\it Herschel}/PACS and SPIRE} 
\software{GILDAS}


{}


\end{document}